\documentclass[prb,amsfonts,amssymb,floats,twocolumn,superscriptaddress,aps]{revtex4-2}
\usepackage[utf8]{inputenc}
\usepackage[T1]{fontenc}
\usepackage{amsmath}
\usepackage{float}
\usepackage[caption = false]{subfig}
\usepackage{color}
\usepackage{braket}
\usepackage{bm}
\usepackage{dsfont}
\usepackage{adjustbox}
\usepackage{graphicx}
\usepackage{enumitem}
\usepackage{bbm}
\usepackage{tikz}
\usetikzlibrary{patterns}

\usepackage[colorlinks=true]{hyperref}
\hypersetup{linkcolor=blue,citecolor=blue,urlcolor=blue}

\renewcommand{\Re}{\operatorname{Re}}
\renewcommand{\Im}{\operatorname{Im}}

\newcommand{\iu}{\mathrm{i}} 
\newcommand{\eu}{\mathrm{e}} 
\newcommand{\du}{\mathrm{d}} 
\newcommand{\hc}{\mathrm{h.c.}} 

\newcommand{\bvec}[1]{{\bm{#1}}}

\newcommand{\Ztwo}{\mathbb{Z}_2}
\newcommand{\Uone}{\mathrm{U(1)}}

\newcommand{\SO}{\mathrm{SO}}

\newcommand{\Ch}{C} 
\newcommand{\ChD}{C^{(\mathrm{D})}} 

\DeclareMathOperator{\tr}{tr}
\DeclareMathOperator{\diag}{diag}

\graphicspath{{./}{figs/}}

\begin{document}

\title{Flux crystals, Majorana metals, and flat bands in exactly solvable spin-orbital liquids}

\author{Sreejith Chulliparambil}
\affiliation{Institut f\"ur Theoretische Physik and W\"urzburg-Dresden Cluster of Excellence ct.qmat, Technische Universit\"at Dresden, 01062 Dresden, Germany}
\affiliation{Max-Planck-Institut f{\"u}r Physik komplexer Systeme, N{\"o}thnitzer Stra{\ss}e 38, 01187 Dresden, Germany}
\author{Lukas Janssen}
\affiliation{Institut f\"ur Theoretische Physik and W\"urzburg-Dresden Cluster of Excellence ct.qmat, Technische Universit\"at Dresden, 01062 Dresden, Germany}
\author{Matthias Vojta}
\affiliation{Institut f\"ur Theoretische Physik and W\"urzburg-Dresden Cluster of Excellence ct.qmat, Technische Universit\"at Dresden, 01062 Dresden, Germany}
\author{Hong-Hao Tu}
\affiliation{Institut f\"ur Theoretische Physik and W\"urzburg-Dresden Cluster of Excellence ct.qmat, Technische Universit\"at Dresden, 01062 Dresden, Germany}
\author{Urban F. P. Seifert}
\affiliation{Institut f\"ur Theoretische Physik and W\"urzburg-Dresden Cluster of Excellence ct.qmat, Technische Universit\"at Dresden, 01062 Dresden, Germany}
\affiliation{Univ Lyon, ENS de Lyon, Univ Claude Bernard, CNRS, Laboratoire de Physique, 69342 Lyon, France}


\date{\today}

\begin{abstract}
Spin-orbital liquids are quantum disordered states in systems with entangled spin and orbital degrees of freedom. 
We study exactly solvable spin-orbital models in two dimensions with selected Heisenberg-, Kitaev-, and $\Gamma$-type interactions, as well as external magnetic fields.
These models realize a variety of spin-orbital-liquid phases featuring dispersing Majorana fermions with Fermi surfaces, nodal Dirac or quadratic band touching points, or full gaps.
In particular, we show that Zeeman magnetic fields can stabilize nontrivial flux patterns and induce metamagnetic transitions between states with different topological character.
Solvable nearest-neighbor biquadratic spin-orbital perturbations can be tuned to stabilize zero-energy flat bands.
We discuss in detail the examples of $\SO(2)$- and $\SO(3)$-symmetric spin-orbital models on the square and honeycomb lattices, and use group-theoretical arguments to generalize to $\SO(\nu)$-symmetric models with arbitrary integer $\nu > 1$.
These results extend the list of exactly solvable models with spin-orbital-liquid ground states and highlight the intriguing general features of such exotic phases. Our models are thus excellent starting points for more realistic modellings of candidate materials.
\end{abstract}

\maketitle


\section{Introduction}

Quantum spin liquids \cite{savary16} are fascinating phases of matter in which strong fluctuations stabilize highly non-trivial ``quantum-disordered'' ground states. They feature long-range entanglement and fractionalized excitations, such as emergent fermions and deconfined gauge fields.
Such ground states are expected, for instance, in systems of antiferromagnetically coupled spin-$1/2$ local moments with Heisenberg spin-rotational symmetry on geometrically frustrated lattices.
While several candidate materials in this regard are available experimentally, only few reliable theoretical results for the relevant models exist and one often has to resort to computationally intensive numerical approaches.

A rare example for an \emph{exactly solvable} model realizing a quantum spin liquid was introduced by Kitaev \cite{kitaev06}. It consists of spins-$1/2$ on a honeycomb lattice with bond-dependent exchange interactions, which thus break the Heisenberg symmetry and frustrate the system.
The exact solution yields gapless itinerant Majorana fermions coupled to a static $\Ztwo$ gauge field. Upon opening up a topologically nontrivial gap, the superselection sectors of the system are given by non-Abelian anyons.
Remarkably, it was later realized that bond-dependent exchange interactions of the Kitaev type naturally occur in transition metal oxides with strong spin-orbit coupling \cite{jackeli09}, paving the way to the experimental study of the so-called Kitaev materials~\cite{trebst2017,janssen2019}.

While the Kitaev model assumes a single spin-$1/2$ degree of freedom per site, systems with interacting spin \emph{and} orbital degrees of freedom have found renewed interest in recent years.
First studied in the context of transition metal oxides with doubly-degenerate $e_g$ orbitals, relevant Kugel-Khomskii models \cite{kk77,kk82,khaliullin2005} have lately been applied to $4d$ and $5d$ systems, in which strong spin-orbit coupling leads to interacting insulators with effective $j_\mathrm{eff}=3/2$ moments \cite{balents10,natori16,jackeli18}, to iron pnictides \cite{kruger2009}, as well as to correlated metallic, insulating and superconducting states, and corresponding transitions, observed in twisted bi- and trilayer structures~\cite{Xu2018,Venderbos2018,Yuan2018,Zhang2019,Classen2019,Schrade2019,xu19,Zhang2020}.

Two key considerations make the search for quantum-disordered phases in spin-orbital systems particularly promising \cite{Ishihara97,Feiner97,Li98,Khaliullin00,Vernay04,Wang09}:
On one hand \cite{savary16}, some spin-orbital models may have $\mathrm{SU}(4)$-symmetric points in their parameter space \cite{kk82,Li98,van2000,van2001,jackeli18}.
It is expected that near such high-symmetry points, quantum fluctuations become enhanced, as magnetic order in generalized Heisenberg antiferromagnets with $\mathrm{SU}(N)$ or $\mathrm{Sp}(N)$ symmetry has been shown to become increasingly unstable upon enlarging the symmetry group, even on unfrustrated lattices either to spin liquid or bond-ordered states \cite{sachdev91,hermele09}, as found also e.g. in continuous-$N$ quantum Monte Carlo studies of the square lattice Heisenberg model \cite{Beach09}.
Indeed, numerical studies \cite{corboz12,natori19} suggest that the $\mathrm{SU}(4)$-symmetric Kugel-Khomskii model on the honeycomb lattice hosts an algebraic spin-orbital liquid, which may explain the disordered ground state in the spin-orbital system $\mathrm{Ba}_{3}\mathrm{CuSb}_{2}\mathrm{O}_{9}$ observed experimentally \cite{nakatsuji2012}.
On the other hand, the anisotropic spatial extent of $d$ orbitals often implies that orbital-orbital interactions in the degenerate subspace are inherently frustrated \cite{khomskii03,balents10}.
Indeed, the Kitaev model is understood to belong to a large class of ``compass'' models \cite{nussinov15} with bond-dependent orbital-orbital interactions, first discussed by Khomskii and Kugel \cite{kk73}.
One may thus expect that appropriate systems with frustrated inter-orbital interactions may host quantum-disordered ground states.
A prominent example is the double perovskite $\mathrm{Ba}_{2}\mathrm{YMoO}_{6}$, which has effective $j_\mathrm{eff}=3/2$ moments as a result of degenerate $t_{2g}$ orbitals and spin-orbit coupling, and does not order down to low temperatures \cite{vries2010,Aharen2010,Carlo2011,vries2013}. Theoretical analyses show that frustration due to certain bond-dependent interactions indeed destabilizes order~\cite{balents10}, and put forward (not exactly solvable) Kitaev-type spin liquids as candidate ground states \cite{natori16}.

In this work, we study models for spin-orbital liquids defined on square and honeycomb lattices. These models belong to a family of generalized Kitaev models that can be solved exactly \cite{CSVJT20}.
They feature bond-dependent biquadratic spin-orbital interactions, while possessing global $\SO(\nu)$ spin-rotational symmetry with integer $\nu > 1$.
We show that these models can be amended by a large number of realistic (and not necessarily small) microscopic perturbations, under which the system remains exactly solvable in terms of dispersing Majorana fermions and static $\Ztwo$ gauge fields.
These perturbations include onsite Zeeman magnetic fields as well as further bond-dependent Kitaev- and off-diagonal $\Gamma$-type exchange interactions, which break the global spin-rotational symmetry.
In particular we find that, as a function of magnetic field, a series of metamagnetic transitions occur due to changes in the ground-state flux configuration.
Various states with metallic or semimetallic Majorana Fermi surfaces, such as ``Majorana metals''~\cite{trebst14} and Dirac or quadratic-band-touching
semimetals, as well as fully gapped states, are stabilized during the magnetization process.
The different states and the corresponding transitions can be characterized via the topology of the pertinent Majorana Bloch wave functions.
This implies that upon adding a small time-reversal symmetry-breaking three-body interaction~\cite{kitaev06}, the semimetallic states acquire a topologically non-trivial gap:
At zero external magnetic field, the $\SO(\nu)$-symmetric model is characterized by a nontrivial Chern number $\Ch = \nu$.
For even (odd) $\nu$, it hosts Abelian (non-Abelian) anyonic excitations with topological spin $\theta = \frac{\pi}{8}(\nu \bmod 16)$ \cite{CSVJT20}.
Upon increasing the field strength, we find that the $\nu =2$ model on the square lattice features a second $\Ch = 2$ state at finite fields, while in the $\nu=3$ model on the honeycomb lattice, a field-induced Abelian $\Ch=4$ phase is encountered.
This is in resemblance to the higher-Chern-number states found in the original Kitaev model perturbed by additional exchange interactions and magnetic fields \cite{batista19, halasz20, ZXLiu2019,QHWang2019}.
Beyond a certain field strength, the $\nu=2$ model exhibits a macroscopic ground-state degeneracy with a trivial fermionic spectrum,
while the $\nu=3$ model features a single dispersing Majorana fermion coupled to static $\Ztwo$ fluxes, corresponding to a Chern number $\Ch=1$.
We furthermore show how some of our explicit results for $\nu = 2$ and $\nu=3$ generalize to the $\SO(\nu)$-symmetric models with $\nu > 3$.
In particular, we find that onsite terms (such as generalized magnetic fields) provide a systematic way to reduce the Chern number $\Ch \to \Ch -2$ in the topologically ordered phases realized in arbitrary-$\nu$ models.

We thus significantly expand the list of exactly solvable spin-orbital models, for which rigorous results can be established \cite{deCarva18,natori20}.
While the exchange interactions in candidate materials contain additional perturbations that spoil the exact solvability of the fine-tuned models considered here, our results can help to highlight the general characteristics of spin-orbital models with bond-dependent interactions, and reveal the exotic properties of quantum spin-orbital liquids.

The rest of the paper is organized as follows.
We discuss the models, relevant symmetries, as well as previously established results in Sec.~\ref{sec: solvable_kitaev}.
Section~\ref{sec:perturbations} contains a classification of perturbations that preserve the solvability of the models.
In Sec.~\ref{sec:field}, we study the effects of a Zeeman magnetic field that couples to the spin degrees of freedom, and we characterize occurring topological transitions.
We discuss selected nearest-neighbor spin-orbital interactions that break the spin-rotational symmetry in Sec.~\ref{sec:NN_int}.
The generalization to $\SO(\nu)$-symmetric models with arbitrary $\nu>1$ is given in Sec.~\ref{sec:generalization}.
Section \ref{sec: conc} concludes the paper.
Technical details on the variational ground-state-flux-sector search are deferred to the Appendix.

\section{Solvable Kitaev spin-orbital liquids} \label{sec: solvable_kitaev}

\subsection{Models and symmetries} \label{sec:modelsymm}

The spin-orbital liquids considered here are the $\nu = 2$ and $\nu = 3$ instances of a family of exactly solvable $\SO(\nu)$-symmetric generalizations of Kitaev's $\Ztwo$ spin liquid recently proposed \cite{CSVJT20}. They are defined on the square and honeycomb lattices, respectively, and feature $\nu$ itinerant Majorana fermions coupled to the same $\Ztwo$ gauge field.
For even $\nu = 2q$ (odd $\nu = 2q+1$) with integer $q \geq 0$, the Hamiltonian reads
\begin{equation}
    \mathcal{H}^{(\nu)}_J  =  -\sum_{\langle ij\rangle_{\gamma}} J_{\gamma}
    \left(\Gamma_{i}^{\gamma}\Gamma_{j}^{\gamma} + \sum_{\beta=\gamma_{\textrm{m}}+1}^{2q+3} \Gamma_{i}^{\gamma\beta} \Gamma_{j}^{\gamma\beta}\right),
\label{eq:h_gamma}
\end{equation}
where $\langle ij \rangle_\gamma$ denotes a nearest-neighbor $\gamma$-type bond on the square (honeycomb) lattice, with $\gamma=1,\dots,\gamma_\mathrm{m}$ and $\gamma_\mathrm{m} = 4$ ($3$) the lattice coordination number. The Gamma matrices $\Gamma^\alpha$, $\alpha = 1, \dots, 2q+3$, form a $2^{q+1}$-dimensional representation of the Clifford algebra, and $\Gamma^{\alpha \beta} = \iu [\Gamma^\alpha, \Gamma^\beta] /2$ for $\alpha < \beta$. 
The Kitaev honeycomb model is recovered for $q=0$ and $\nu=1$, with the usual three Pauli matrices as two-dimensional Gamma-matrix representation, $(\Gamma^\alpha)_{\alpha = 1,2,3}=(\sigma^x,\sigma^y,\sigma^z)$.
The exact solvability of the model relies on representing the Gamma matrices in terms of $2 q+ 4$ Majorana fermions $c,b^\alpha$ as $\Gamma^\alpha = \iu b^\alpha c$~\cite{yzk09,wu2009,ryu2009}, yielding a problem of $\nu$ Majorana fermions dispersing in the background of a static $\Ztwo$ gauge field $u_{ij} = i b_i^\gamma b_j^\gamma$ on a $\langle ij \rangle_\gamma$ link,
\begin{equation}\label{eq:h_gamma_u}
    \tilde{\mathcal{H}}^{(\nu)}_J  =  \sum_{\langle ij\rangle_{\gamma}} J_{\gamma} u_{ij} \left( \iu c_i c_j + \sum_{\beta=\gamma_{\textrm{m}}+1}^{2q+3} \iu b^\beta_i b^\beta_j \right).
\end{equation}
 The representation of the Gamma matrices in terms of Majorana fermions introduces additional unphysical states which can be projected out by demanding the constraint $D_j = \iu^{q+2} b_j^{1} \dots b_j^{2q+3}c_j = -1$ that holds only in the subspace of \emph{physical} states.
Note that $ \tilde{\mathcal{H}}^{(\nu)}$ possesses an $\mathrm{O}(\nu)$ symmetry of rotations of the $\nu$-component spinor $(c_i, b^{\gamma_\mathrm{m}}_i, \dots, b^{2q+3}_i)^\top$.
This global $\mathrm{O}(\nu)$ symmetry is also present in the microscopic Hamiltonian $\mathcal{H}^{(\nu)}$:
The (normal) $\SO(\nu)$ subgroup of $\mathrm{O}(\nu)$ is generated by $\sum_j \Gamma^\alpha_j$ and $\sum_j \Gamma^{\alpha \beta}_j$ with $\alpha,\beta = \gamma_\mathrm{m} + 1, \dots , 2q+3$. More details on the $\SO(\nu)$ symmetry algebra are given in the Supplemental Material to Ref.~\cite{CSVJT20}.
For the second connected component of $\mathrm{O}(\nu) = \mathbb{Z}_2 \ltimes \mathrm{SO}(\nu)$, it is sufficient for us to find a single symmetry operation $\rho \in \mathrm{O}(\nu) / \SO(\nu)$ with $\det \rho = -1$, as all other elements then lie in the orbit of $\rho$ under $\mathrm{SO}(\nu)$.
This $\rho$ is given by a generalization of the global dihedral $\mathrm{D}_2$ spin-rotational symmetry in the $\nu=1$ Kitaev model, and we choose it to act on any Gamma matrix $\Gamma^\alpha$ as $\rho: \Gamma^\alpha \mapsto \Gamma^1 \Gamma^\alpha \Gamma^1$ for $\alpha = 1,\dots,2q+3$, such that $\Gamma^1 \mapsto \Gamma^1$ and $ \Gamma^{\alpha} \mapsto - \Gamma^{\alpha}$ for $\alpha \neq 1$, as well as $\Gamma^{1 \beta} \mapsto - \Gamma^{1 \beta}$ and $\Gamma^{\alpha \beta} \mapsto \Gamma^{\alpha \beta}$ for $1 < \alpha < \beta$. 
It is easily verified that $\rho$ is unitary.
In the Majorana parton basis, this generalized dihedral symmetry acts as $b^1_i \mapsto - b^1_i$ and $c_i \mapsto -c_i$, with the remaining Majoranas $b^\beta$, $\beta = 2, \dots, 2q+3$, being invariant, so that model's fermion parity is not changed by the symmetry transformation.
This implies that also $u_{ij} = \iu b^1_i b^1_j \mapsto u_{ij}$ on $\langle ij \rangle_1$ links, such that the gauge field $u_{ij}$ is invariant under $\rho$ on all links $\langle ij \rangle_\gamma$.
As the transformation acts on the itinerant Majorana fermion spinor as
\begin{multline}
    \rho: \quad (c_i, b^{\gamma_\mathrm{m}}_i, \dots, b^{2q+3}_i)^\top \mapsto \\
    \diag(-1,1, \dots, 1) (c_i, b^{\gamma_\mathrm{m}}_i, \dots, b^{2q+3}_i)^\top,
\end{multline}
we conclude that $\rho$ is orthogonal, $\rho \in \mathrm{O}(\nu)$, and fulfills $\det \rho = -1$, such that $\rho \in \mathrm{O}(\nu) / \SO(\nu)$, as required.
For more details on the exact solution of $\mathcal{H}^{(\nu)}$, we refer to Ref.~\cite{CSVJT20}.

The $\nu = 2$ and $\nu = 3$ models have a four-dimensional local Hilbert space, and the mapping to spin-orbital models is achieved by representing the $4 \times 4$ Gamma matrices as $\Gamma^\alpha = -\sigma^y \otimes \tau^\alpha$ for $\alpha = x,y,z$, $\Gamma^4 = \sigma^x \otimes \mathds{1}$, and $\Gamma^5 = -\sigma^z \otimes \mathds{1}$. 
Here, $(\sigma^x,\sigma^y,\sigma^z)$ and $(\tau^x,\tau^y,\tau^z)$ denote two sets of usual $2\times2$ Pauli matrices and are assumed to act on the spin and orbital degrees of freedom, respectively.
It is convenient to relabel the itinerant Majorana fermions as $b^5 \to c^x, c \to c^y$, and further on the honeycomb lattice $b^4 \to c^z$, so that on the square lattice the $z$ component of the spin operator is expressed in terms of the two itinerant Majorana fermions, $\sigma^z \otimes \mathds{1} = - \iu c^x c^y$.
Further, on the honeycomb lattice, the spin operators are expressed in terms of itinerant Majorana fermions as
\begin{equation} \label{eq:majoranaSO3}
  \sigma^\alpha \otimes \mathds{1} = -\frac{\iu}{2} \epsilon^{\alpha \beta \gamma} c^\beta c^\gamma  \equiv \frac{1}{2} c^\top L^\alpha c
\end{equation}
with the $\SO(3)$ generators $L^\alpha_{\beta \gamma} = - \iu \epsilon^{\alpha \beta \gamma}$ in the fundamental representation, and where we have assumed the summation convention over repeated indices $\alpha, \beta, \gamma \in\{x,y,z\}$.
Note that Eq.~\eqref{eq:majoranaSO3} has previously been employed in parton decompositions of spin-$1/2$ systems \cite{tsvelik92,shastrysen97,biswas11}, in which case the local Hilbert space is enlarged by redundant states. 
In the spin-orbital systems we study here, redundant states are projected out in the exact solution by demanding $\Ztwo$ gauge invariance of physical states and  spectrum \cite{kitaev06,CSVJT20}. Operators that act trivially in the spin sector and non-trivially in the orbital sector  involve only ``gauge'' Majoranas when mapped to the Majorana representation,
\begin{equation} \label{eq:tauOps}
   \mathds{1} \otimes \tau^{\alpha} = - \frac{1}{2}\epsilon_{\alpha \beta \gamma }\Gamma^{\beta \gamma} = -\frac{1}{2}\epsilon_{\alpha \beta \gamma }ib^{\beta}b^{\gamma},
\end{equation}
where indices are defined analogous to Eq.~\eqref{eq:majoranaSO3}.
\begin{figure}[!tb]
\includegraphics[width=.95\columnwidth,clip]{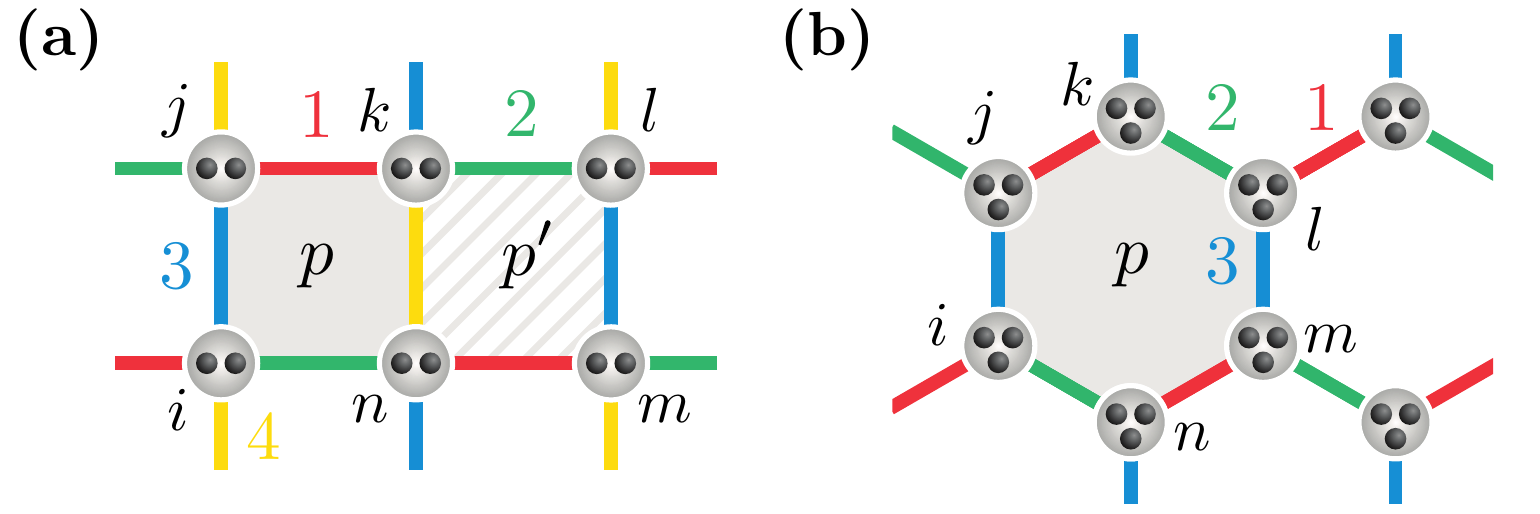}
\caption{
Illustration of square- and honeycomb-lattice Kitaev spin-orbital models. (a) In the $\nu = 2$ model, there are two itinerant Majorana fermions per site. The bond-dependent exchange interactions double the elementary unit cell of the square lattice, and thus there are two inequivalent elementary plaquettes $p$ (shaded) and $p'$ (hatched). (b) In the $\nu=3$ model on the honeycomb lattice, there are three itinerant Majorana fermions per site, and there is only one type of elementary plaquette $p$.
}
\label{fig:lattices}
\end{figure}

\subsubsection{$\nu=2$ model on the square lattice} \label{sec:modelsymm_SQ}

On the square lattice, one obtains in the spin-orbital basis
\begin{equation} \label{eq:h_sq}
  \mathcal{H}^{(2)}_J= - \sum_{\langle i j \rangle_\gamma} J_\gamma \left( \sigma^x_i \sigma^x_j + \sigma^y_i \sigma^y_j \right) \otimes \tau^\gamma_i \tau^\gamma_j,
\end{equation}
where $\gamma = 1,2,3,4$ denotes the four inequivalent bonds in a two-site unit cell and $(\tau^\gamma) = \left(\tau^x,\tau^y,\tau^z,\mathds{1} \right)$.
In the following, we will restrict to isotropic $J_\gamma \equiv J > 0$ for simplicity.
It is straightforward to see that the Hamiltonian~\eqref{eq:h_sq} possesses a global $\SO(2)$ symmetry of spin rotations about the $z$ axis.
Due to the bond-dependent nature of the Kitaev-type orbital interactions, lattice symmetry operations also act simultaneously on the orbital degrees of freedom.
Specifically, we consider perturbations that preserve the following symmetries of $\mathcal H^{(2)}$:
\begin{enumerate}[label=(\arabic*)]
    \item A fourfold rotational symmetry about the center of a plaquette, which also acts on the orbitals as $\mathds{1} \mapsto \tau^x \mapsto \tau^y \mapsto \tau^z \mapsto \mathds{1}$.
    \item A reflection across an axis perpendicular to the $3$- and $4$-bonds, which maps $\tau^x \mapsto -\tau^y$, $\tau^y \mapsto -\tau^x$ and $\tau^z \mapsto - \tau^z$ and $\mathds{1} \mapsto \mathds{1}$.
\end{enumerate}
We further allow the $\SO(2)$ spin-rotational symmetry to be broken, but assume that spin interactions along each bond $\langle ij \rangle_\gamma$ are invariant under
\begin{enumerate}[label=(\arabic*)]
    \setcounter{enumi}{2}
    \item Inversion with $\sigma^\alpha_{i} \leftrightarrow \sigma^\alpha_j$ for all $\alpha = x,y,z$.
    \item Rotations of $\pi/2$ about the $z$ axis, which maps $\sigma^x \mapsto -\sigma^y$, $\sigma^y \mapsto  \sigma^x$ and $\sigma^z \mapsto \sigma^z$.
\end{enumerate}
Importantly, we note that the model defined in Eq.~\eqref{eq:h_sq} possesses an extensive number of conserved quantities given by the plaquette operators
\begin{equation}
  W_p = - \Gamma^{23}_i \Gamma^{31}_j \Gamma^{14}_k \Gamma^{42}_n \quad \text{and} \quad W_{p'} = - \Gamma^{42}_k \Gamma^{23}_l \Gamma^{31}_m \Gamma^{14}_{n}
\end{equation}
on the two inequivalent elementary plaquettes of the square lattice, as shown in Fig.~\ref{fig:lattices}(a).
Rewriting the Gamma matrices in terms of spin and orbital degrees of freedom yields
\begin{subequations}\label{eq:W_sq}
  \begin{align}
    W_p &= \sigma_k^z \sigma_n^z \otimes \tau^x_i \tau^y_j \tau^x_k \tau^y_n, \label{eq:W_p_a} \\
    W_{p'} &= \sigma^z_k \sigma^z_n \otimes \tau^y_k \tau^x_l \tau^y_m \tau^x_n. \label{eq:W_p_b}
  \end{align}
\end{subequations}
Note that the orbital components of the above operators are equivalent to the plaquette operators of Wen's exactly soluble model for $\Ztwo$ gauge theory \cite{wen03}.

\subsubsection{$\nu=3$ model on the honeycomb lattice} \label{sec:modelsymm_HC}

The $\nu = 3$ model on the honeycomb lattice, rewritten in the spin-orbital basis, reads
\begin{equation}
  \mathcal{H}^{(3)}_J= - \sum_{\langle ij \rangle_\gamma} J_\gamma \left(\vec \sigma_i \cdot \vec \sigma_j \right) \otimes \tau^\gamma_i \tau^\gamma_j,
\end{equation}
where $\gamma = x,y,z$, $\vec \sigma = (\sigma^x, \sigma^y, \sigma^z)$ and we again assume isotropic $J_\gamma \equiv J>0$.
In this spin-orbital basis, the global $\SO(3)$ symmetry hence corresponds to an $\SO(3)$ spin-rotational symmetry.
As in the square-lattice model, the highly anisotropic interactions in the orbital sector imply that lattice symmetry operations also act on the orbital degrees of freedom.
We take the point group $\mathrm{C}_{6v} \simeq \mathrm{D}_{3d}$ to be generated by
\begin{enumerate}[label=(\arabic*)]
    \item A sixfold rotational symmetry $C_6$ about the center of a hexagonal plaquette, which also maps the components of orbital operators $(x,y,z) \mapsto (y,z,x)$. We further assume that the spin degrees of freedom are also coupled to the lattice and thus also transform under $C_6$ as $(x,y,z) \mapsto (y,z,x)$.
    \item A reflection symmetry $\sigma$ across an axis perpendicular to the $z$ bonds, which acts on \emph{both} orbital and spin operators as $(x,y,z) \mapsto (y,x,-z)$.
\end{enumerate}
Note that our assumption that the spin degrees of freedom are coupled to the lattice and thus also transform under $C_6$ is due to the convenient fact that in the $\nu = 3$ model, there are \emph{three} Pauli matrices (spin components) which may be distributed on the \emph{three} links of the honeycomb lattice. 
Note that this is different from the situation of the $\nu =2$ model, but an analogous property can be found in the $\nu=4$ model on the square lattice, for which the natural four-dimensional representation of $\SO(4)$ can be placed on the four distinct link types.

The conserved plaquette operators on the honeycomb lattice, which ensure that the fluxes are static read
\begin{equation} \label{eq:W_hc}
  W_p = \mathds{1} \otimes \tau^x_i \tau^y_j \tau^z_k \tau^x_l \tau^y_m \tau^z_n,
\end{equation}
see Fig.~\ref{fig:lattices}(b). This makes explicit that the flux operators only involve orbital degrees of freedom, as previously noted \cite{yaolee11,natori20}.

\subsection{Review of previous results for Kitaev-type spin-orbital liquids}

We note that our $\nu =2$ model has been previously studied in the basis of $j_\mathrm{eff}=3/2$ operators by Yao, Zhang, and Kivelson \cite{yzk09}, as well as Nakai, Ryu, and Furusaki \cite{nakai2012}. The $\nu =3$ model on a decorated honeycomb lattice, leading to spontaneous time-reversal symmetry breaking, was studied by Yao and Lee \cite{yaolee11}.
Few subsequent works have utilized the exact solvability of these models and established rigorous results, which we review here. These results can straightforwardly be extended to the perturbed models we discuss below.

In Ref.~\cite{deCarva18}, the $\nu=3$ model was studied on the honeycomb lattices with zigzag edge terminations. Flat edge states and algebraically decaying spin-spin correlations $\langle \vec \sigma(\bvec x) \cdot \vec \sigma(0) \rangle \sim 1/|\bvec x|^4$ were found to generate a non-local edge magnetization upon applying local magnetic fields.
Since the model possesses an $\SO(3)$ spin-rotational symmetry, the Majorana excitations carry spin, in contrast to the $\nu=1$ Kitaev model. This might allow one to probe spin transport properties in heterostructures consisting of spin-orbital liquids sandwiched between two metallic plates.
It is found that the spin current shows a power-law dependence on the applied spin potential $V = \mu_\uparrow - \mu_\downarrow$, with the exponent allowing the distinction of semimetallic nodal structures from the fully gapped spectrum of the chiral Kitaev spin-orbital liquid.
Furthermore, a longitudinal spin Seebeck effect is predicted, i.e., a finite spin current that is mainly due to edge states and being driven by a temperature gradient between the two edges in the presence of a magnetic field.

In a recent work, Natori and Knolle studied the dynamic and spectroscopic properties of the $\nu=3$ model on the honeycomb lattice \cite{natori20}, utilizing a mapping to quantum quenches previously developed for the $\nu=1$ Kitaev model \cite{knolle14}.
They find that the dynamic structure factor consists of two contributions: The first is given by the dynamic spin-spin correlation function $\langle \sigma^\alpha_i(t) \sigma^\beta_j(0) \rangle$, which maps onto density-density correlation function of the itinerant Majorana fermions and has an algebraic (exponential) decay in the gapless (gapped) phases.
The second contribution is given by spin-orbital correlation functions, which involve the excitations of $\Ztwo$ gauge fluxes and thus has an exponential form, with the corresponding gap being three times as large as in the $\nu=1$ Kitaev model.
While the structure factor is probed in neutron scattering experiments, the authors suggest that the spin dynamics in the relevant $4d^1$ and $5d^1$ Mott insulators may be separately probed using resonant inelastic X-ray scattering~\cite{natori20}.

\section{Solvable perturbations}\label{sec:perturbations}

It is straightforward to see from the representation of the plaquette operators in the spin-orbital basis in Eqs.~\eqref{eq:W_sq} and \eqref{eq:W_hc} that, in addition to the pure Kitaev interactions, there are many possible perturbations that commute with the flux operators and thus keep the gauge field static.
For the $\nu=3$ model on the honeycomb lattice, this in particular applies to interactions that only couple to the spin degrees of freedom.
For the $\nu=2$ model on the square lattice, spin interactions that commute with $\sigma^z_i \sigma^z_j$ on $\langle i j \rangle_4$ links maintain this property of the unperturbed model.
In this section, we classify such solvable (not necessarily small) perturbations with respect to their symmetry properties and the range of interaction.

We emphasize that the majority of perturbations of this form generically lead to interactions among the itinerant Majorana degrees of freedom. In the strongly-interacting regime, such perturbations may induce fractionalized quantum critical points between the disordered spin-orbital-liquid and partially-ordered phases~\cite{seifert20}.
By contrast, here we consider perturbations that are quadratic in the dispersing Majoranas and thus preserve the exact solvability of the model.
As different species of itinerant Majorana fermions $c^\alpha_i$ couple identically to the gauge field $u_{ij}$, any $\Ztwo$ gauge transformation $u_{ij} \mapsto s_i u_{ij} s_j$ with $s_i = \pm 1$ acts identically on all flavors $\alpha$, $c^\alpha_i \mapsto s_i c^\alpha_i$.
Gauge invariance then demands that any solvable perturbation, involving a bilinear of itinerant Majorana fermions at sites $i$ and $i+l$, will be of the form 
\begin{equation} \label{eq:h_pr}
  \tilde{\mathcal{H}}' \sim f^{\alpha \beta}_{i,i+l} \iu c^\alpha_i \Bigg[ \prod_{\langle jk \rangle \in \mathcal L} u_{jk} \Bigg] c^\beta_{i+l},
\end{equation}
where $\mathcal L$ denotes a path through the lattice connecting sites $i$ and $i+l$.
The couplings $f^{\alpha \beta}_{i,i+l}$ depend on the microscopic nature of the interactions and can be constrained by symmetry.
Here, we focus mainly on onsite ($l=0$) and nearest-neighbor $(l=1$) perturbations.
Perturbations that lead to longer-ranged hopping of a single Majorana flavor (on the honeycomb lattice) have been discussed previously by Kitaev ($l=2$) ~\cite{kitaev06}
and Zhang \textit{et al.} ($l=3$) \cite{batista19} for the original Kitaev model and may be easily generalized to the spin-orbital models discussed here.

Once the ground state of the gauge field is known (labelled in a gauge-invariant manner by a configuration of the plaquette operators $W_p = \pm 1$), the full Hamiltonian $\tilde{\mathcal{H}} + \tilde{\mathcal{H}}'$ describes a problem of noninteracting hopping fermions and can be diagonalized straightforwardly.
We emphasize that at finite perturbation strengths, the system in general does not possess reflection symmetry, and thus Lieb's theorem \cite{lieb94} no longer holds, implying that the $\pi$-flux (flux-free) phase may be no longer the ground state of the square-lattice (honeycomb-lattice) model.
Therefore, to find the optimal configuration of the $\Ztwo$ gauge field for a given parameter set, we diagonalize the respective fermionic hopping problems on finite-size lattices of $48 \times 48$ unit cells in the background of a variety of flux configurations in order to uncover the ground-state flux pattern.
The flux sectors considered are shown together with an exemplary corresponding gauge-field configuration $\{ u_{ij} \}$ in the Appendix.

We have verified that the obtained phase boundaries vary only slightly upon further increasing the system sizes up to $60 \times 60$ unit cells for selected points in parameter space.

\subsection{Onsite terms: Magnetic fields}

We first discuss solvable onsite perturbations to the spin-orbital liquids introduced above.
As shown below, these terms correspond to Zeeman magnetic fields that couple only to the spin degrees of freedom.
Such a spin-only coupling has previously been shown to result from a strong-coupling expansion of a Hubbard model with twofold orbital degeneracy~\cite{kk77}. 
We note that in general, spin-orbital coupling will lead to additional terms that couple the orbital degrees of freedom to the external field. However, such terms will involve single $\tau$ operators that do not commute with the plaquette operators $W_p$ and thus lead to dynamics of the fluxes, spoiling the exact solvability of the model. We therefore leave an analysis of the effects of this orbital coupling for further studies.

\subsubsection{$\nu=2$ model on the square lattice} 

On the square lattice, we consider a Zeeman field in the $z$ direction, coupling to the spin degrees of freedom as
\begin{equation}
  \mathcal{H}^{(2)}_h = - h^z \sum_i \sigma^z_i \otimes \mathds{1}.
\end{equation}
$\mathcal{H}^{(2)}_h$ indeed commutes with the flux operators in Eq.~\eqref{eq:W_sq} and is seen to be quadratic in the itinerant Majorana fermions upon writing $\sigma^z \otimes \mathds{1} = -\Gamma^5 = - \iu c^x c^y$, yielding
\begin{equation}
  \tilde{\mathcal{H}}^{(2)}_h = h^z  \sum_i \iu c^x_i c^y_i,
\end{equation}
such that a finite $h^z$ hybridizes the two Majorana flavors.
The comparison with Eq.~\eqref{eq:h_pr} shows that no further solvable onsite terms exist.

Note that the $\nu =2$ model in the presence of a finite $h^z$ field has been previously studied in Ref.~\cite{yzk09}.
However, their study primarily covers the two limiting cases of small $h^z \ll J$ and large fields $h^z \gg J$.
In the former case, the $\pi$-flux gauge-field configuration was argued to be protected by a finite flux gap, such that a small field only alters the dispersion of itinerant Majorana fermions, while in the latter limit of large fields \emph{and} anisotropy in the Kitaev couplings $J$, the model was mapped onto a pure $\Ztwo$ gauge theory with a $\pi$ flux in every plaquette.
These results led the authors to conjecture that these two limits are adiabatically connected.
We show below, however, that there are first-order transitions at intermediate field strengths associated with other flux patterns having lower energy than the $\pi$-flux configuration.
Further, we clarify that at $h^z \gg J$ and isotropic couplings, the system possesses an extensive degeneracy making it unstable towards confinement.

\subsubsection{$\nu=3$ model on the honeycomb lattice}

Similar to the square-lattice case, onsite terms quadratic in the itinerant Majorana fermions are generated by coupling the spin operators to a magnetic field,
\begin{equation}
  \mathcal{H}^{(3)}_h = - \vec h \cdot \sum_i \vec \sigma_i \otimes \mathds{1}.
\end{equation}
Mapping the spin operators to Majorana fermions, it is straightforward to see that the above term exhausts all gauge-invariant quadratic onsite terms,
\begin{equation}
  \tilde{\mathcal{H}}^{(3)}_h = \sum_i \left( h^x \iu c^y_i c^z_i + h^y \iu c^z_i c^x_i + h^z \iu c^x_i c^y_i \right).   
\end{equation}

\subsection{Nearest-neighbor interactions}  \label{sec:nnn-perturb}

Nearest-neighbor terms $\mathcal{H}'$ in Eq.~\eqref{eq:h_pr} on $\langle ij \rangle_\gamma$ bonds generically result from products of the spin-orbital operators $\sigma^y \otimes \tau^\gamma \equiv -\Gamma^\gamma = -\iu b^\gamma c^y$,  $\sigma^x \otimes \tau^\gamma = -\Gamma^{\gamma 5} = -\iu b^\gamma c^x$, and, on the honeycomb lattice, $\sigma^z \otimes \tau^\gamma \equiv - \Gamma^{\gamma 4} = -\iu b^\gamma c^z$, on two adjacent sites, such that $\iu b
^\gamma_i b^\gamma_j = u_{ij}$ forms the $\Ztwo$ gauge field.
It thus becomes clear that \emph{any} nearest-neighbor spin interaction paired with a bond-dependent orbital Ising interaction preserves the conservation of the flux operator and is furthermore quadratic in the itinerant fermions.
A distinctive feature of these additional interactions is that they break the system's $\SO(\nu)$ spin-rotational symmetry and may thus be useful to study properties of spin-orbital liquids away from highly symmetric points.

\subsubsection{$\nu = 2$ model on the square lattice} 

The symmetries given in Sec.~\ref{sec:modelsymm_SQ} constrain solvable nearest-neighbor perturbations to be given by
\begin{equation} \label{eq:bargamma-sq}
    \mathcal{H}_{\bar\Gamma}^{(2)} = \bar\Gamma \sum_{\langle ij \rangle_\gamma} \left( \sigma^x_i \sigma^y_j + \sigma^y_i \sigma^x_j \right) \otimes \tau^\gamma_i \tau^\gamma_j,
\end{equation}
which retains a global discrete spin-rotational symmetry.

\subsubsection{$\nu = 3$ model on the honeycomb lattice} 

Within the symmetry group given in Sec.~\ref{sec:modelsymm_HC}, the set of solvable nearest-neighbor spin-orbital perturbations to $\mathcal{H}^{(3)}_J$ can be constrained to be of the form
\begin{align} \label{eq:hkgg}
    \mathcal{H}_{K\Gamma\Gamma'}^{(3)} = &\sum_{\langle ij \rangle_{\gamma}} \Big[-K \sigma^\gamma_i \sigma^\gamma_j + \Gamma \left(\sigma^\alpha_i \sigma^\beta_j + \sigma^\beta_i \sigma^\alpha_j \right) \nonumber\\ &+ \Gamma'\left(\sigma^\gamma_i \sigma^\alpha_j + \sigma^\alpha_i \sigma^\gamma_j + \sigma^\gamma_i \sigma^\beta_j + \sigma^\beta_i \sigma^\gamma_j \right) \Big] \otimes \tau^{\gamma}_i \tau^{\gamma}_j,
\end{align}
where $(\alpha,\beta,\gamma) = (y,z,x)$, $(z,x,y)$, and $(x,y,z)$ on $x$, $y$, and $z$ bonds, respectively.
The full Hamiltonian $\mathcal{H}^{(3)}_J+ \mathcal{H}_{K \Gamma\Gamma'}^{(3)}$ thus corresponds to a Kitaev-Heisenberg-Gamma-Gamma' model in the spin sector coupled to an orbital Kitaev model. It is exactly solvable at every point in parameter space, thus significantly extending the list of rare examples of exactly-soluble spin-orbital liquids.

\section{Field-induced phases} \label{sec:field}

In the following, we discuss the phases obtained in the $\nu = 2$ and $\nu =3$ models upon coupling the spin degrees of freedom to external magnetic fields. We first discuss the ground-state flux sectors and magnetization, and then characterize the occurring phases according to the topology of the free-fermion wave functions in the respective lowest-energy flux sectors.

\subsection{Flux patterns}

\subsubsection{$\nu=2$ model on the square lattice} \label{sec:pd_square}

\begin{figure}[!tb]
\includegraphics[width=.9\columnwidth,clip]{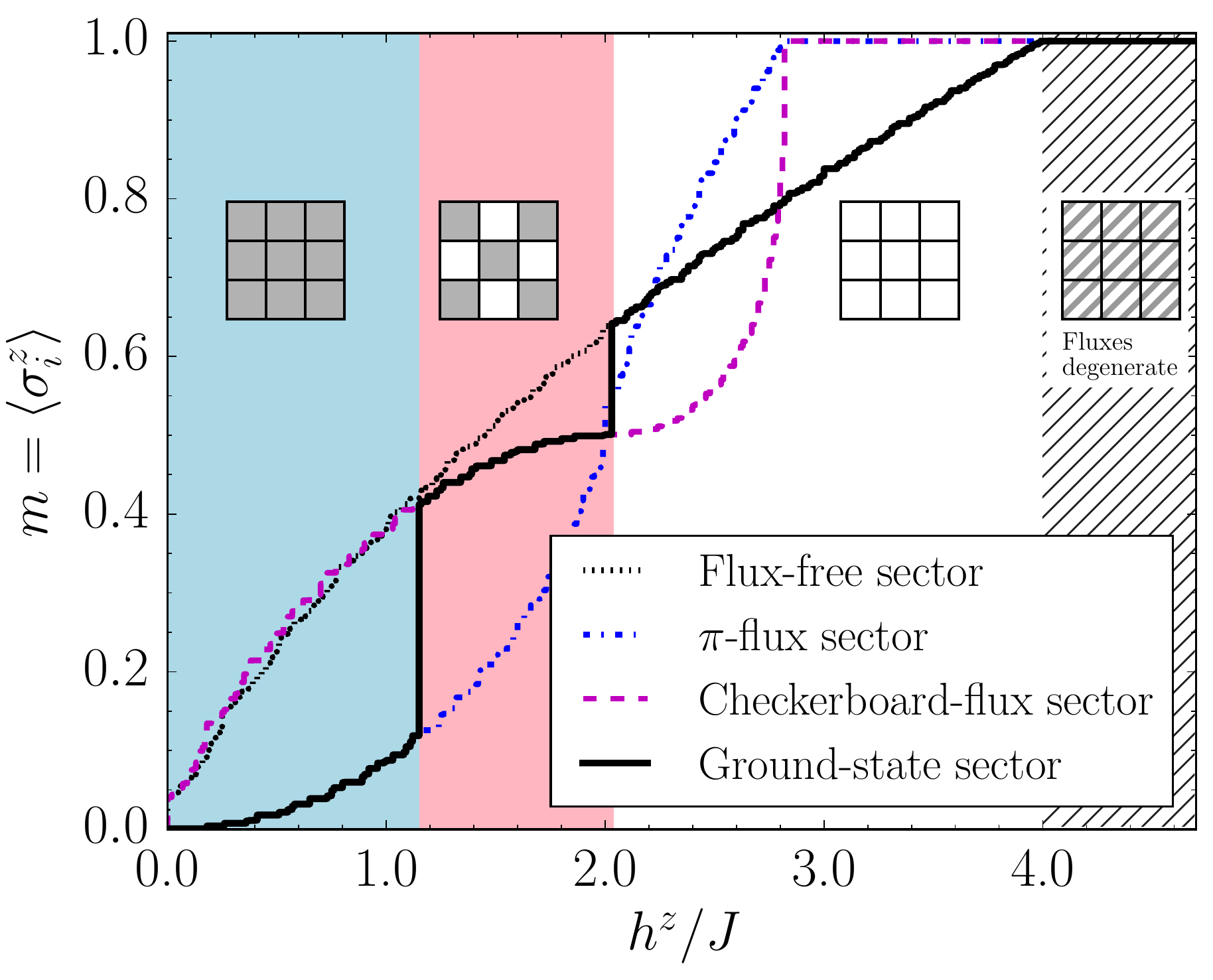}
\caption{
Ground-state flux configuration and magnetization $m$ per site for the square-lattice spin-orbital liquid as a function of magnetic field along the $z$ axis. The continuous black line corresponds to the magnetization in the respective lowest-energy flux sector, while the dashed lines indicate the magnetizations in the three indicated flux sectors.
Here, we have used a lattice with $48 \times 48$ unit cells and employed a small finite temperature $T = 0.001 J$ for numerical stability. Small jumps in the magnetization are due to finite-size effects.
The insets indicate the corresponding flux patterns, where gray (white) squares correspond to $\pi$-flux ($0$-flux) plaquettes.
At low fields, the $\pi$-flux state is stable (blue area). Intermediate fields $1.1 \lesssim h^z/J \lesssim 2.0$ induce a checkerboard flux pattern (violet area), while a flux-free state is stabilized for $2.0 \lesssim h^z/J < 4$ (white area).
For $h^z/J > 4$, all flux configurations become degenerate (hatched area).}
\label{fig:sqMag}
\end{figure}

The ground-state flux configuration of the $\nu=2$ model in a Zeeman magnetic field is displayed in Fig.~\ref{fig:sqMag}, together with the longitudinal magnetization per site $m^z = \langle \sigma^z_i \rangle$.
For a given flux configuration, the spectra of $\mathcal{H}^{(2)}_J+ \mathcal{H}^{(2)}_h$ at finite $h^z$ can be studied easily by noting that the two Majorana fermions $c^x_i$ and $c^y_i$ at each site can be combined into a complex fermion $f_i$ as $f_i =  \left(c^x_i + \iu c^y_i \right) / 2$, such that $\iu c^x_i c^y_i = 2 f^\dagger_i f_i -1$.
It is straightforward to see that the $\SO(2)$ symmetry of mixing $c^x$ and $c^y$ then becomes the $\Uone$ phase-rotational symmetry $f_i \mapsto \eu^{\iu \varphi} f_i$, $\phi \in [0, 2\pi)$.
In this formulation, the Hamiltonian reads
\begin{align} \label{eq:square_f}
  \tilde{\mathcal{H}}^{(2)}_J+ \tilde{\mathcal{H}}^{(2)}_h & = 
  J \sum_{\langle i j \rangle} u_{ij} \left( 2 \iu f_i^\dagger f_j + \hc \right)
  \nonumber \\*  & \quad 
  + h^z \sum_i \left( 2 f_i^\dagger f_i - 1 \right),
\end{align}
where h.c.\ stands for Hermitian conjugation, such that the magnetic field $h^z$ takes the role of a chemical potential for the spinless complex fermions hopping in the background of the static $\Ztwo$ gauge field.

By Lieb's theorem~\cite{lieb94}, the ground state at $h^z = 0$ lives in the $\pi$-flux sector with $W_{p} = W_{p'} = -1$ on all plaquettes $p$ and $p'$.
We fix a gauge such that $u_{ij} = +1$ for $i \in A$, $j \in B$ sublattices on $\langle ij \rangle_{\gamma = 1,2,3}$ links and $u_{ij} = -1$ on $\langle ij \rangle_{4}$ links.
Introducing Fourier modes $f_{s,\bvec k} = N^{-1/2} \sum_{i} \eu^{\iu \bvec k \cdot \bvec x_i} f_{s,i}$ on the respective sublattices $s=1,2$
and diagonalizing the resulting $2 \times 2$ Hamiltonian yields the dispersion $\varepsilon_{1,2} (\bvec k) = 2 h^z \pm |g(\bvec k)|$ with $g(\bvec k ) = 2J(1+ \eu^{\iu \bvec k \cdot \bvec n_1} + \eu^{\iu \bvec k \cdot \bvec n_2} - \eu^{\iu \bvec k \cdot (\bvec n_1 + \bvec n_2)})$,
where $\bvec n_{1,2} = (\pm 1,1)$ denote the lattice vectors on the square lattice with a two-site unit cell.
A small magnetic field thus shifts the Fermi level away from the nodal Dirac points, leading to a metallic Majorana Fermi surface and finite magnetization, as shown in Fig.~\ref{fig:sqMag}.

At $h^z \approx 1.1 J$ the system undergoes a first-order transition, due to the checkerboard-flux crystal becoming the lowest-energy flux configuration.
The checkerboard-flux crystal is characterized by alternating $0$-flux and $\pi$-flux plaquettes in a physical unit cell with two sites.
However, due to the projective implementation of the translational symmetry, the unit cell for the Majorana-fermion-hopping problem is enlarged and consists of four sites. This leads to the $f$-fermion spectrum consisting of four bands, with the Fermi level (set by $h^z$) located in the second-lowest band.

Further increasing $h^z$, another first-order phase transition is encountered at $h^z \approx 2.03 J$, with the gauge field now ordering in the flux-free state with $W_p = W_{p'} = +1$ on all plaquettes $p$ and $p'$.
The dispersion in this flux-free background is given by a single band $\varepsilon(\bvec k) = 4 J \left( \cos k_x + \cos k_y \right) + 2 h^z$.
As no further first-order transitions intervene, it is thus clear that for $h^z \geq 4 J$ no fermionic states are occupied, and the spins are now fully polarized, $\langle \sigma_i^z \rangle = +1$, as also visible from Fig.~\ref{fig:sqMag}.
Importantly, the Fermi level lying outside the band, and thus the entire band being completely empty,
implies that a variation of the background gauge field's flux configuration no longer leads to changes in the ground-state energy.
Hence, all flux configurations $\{ W_p = \pm 1\}$ become degenerate for $h^z > 4 J$, and the system possesses an extensive quantum ground state degeneracy scaling as $\sim 2^N$. This originates from the disordered orbital degrees of freedom, with spin sector being fully polarized.
In fact, in the limit $J/h^z \to 0$, the ground state is determined by $\mathcal{H}_h^{(2)}$ as a manifold of spin-polarized orbital-degenerate ground states $\ket{\psi_\uparrow} = \prod_i \ket{\uparrow}_i \otimes \ket{\{\tau_i\}}$. At small, but finite $0< J/h^z \ll 1$, we find that the degeneracy is not lifted in perturbation theory, because the field-polarized state is an eigenstate of $\mathcal{H}^{(2)}_J$ with $\mathcal{H}^{(2)}_J \ket{\psi_\uparrow} = 0$.
The extensive degeneracy may be lifted by a perturbation $\mathcal{H}_\lambda$ that couples to the orbital degrees of freedom, and acts non-trivially on the spin degrees of freedom. 
One may then consider two distinct scenarios:
\begin{enumerate}[label=(\arabic*)]
    \item The perturbation commutes with the plaquette operators, $[\mathcal{H}_\lambda, W_p| = 0$, such that the $\Ztwo$ gauge theory remains in the deconfined phase with a finite flux gap. This scenario was implicitly discussed in Ref.~\cite{yzk09}, where in the limit analogous to our large magnetic fields, introducing anisotropies in the Kitaev couplings, the system gave way to an effective Wen plaquette model~\cite{wen03}, which is equivalent to the toric code~\cite{kitaev03}.
    \item The extensive degeneracy implies that the flux gap vanishes and thus the spin-polarized spin-orbital liquid is unstable to confinement, if one adds small perturbations that spoil the conservation of the plaquette operators, $[\mathcal{H}_\lambda, W_p] \neq 0$. 
    In this confined phase, the effective excitations are usual bosonic modes associated with the resulting ordered state, e.g., magnons.
    For example, one may consider exchange interactions $\sim \vec \tau_i \cdot \vec \tau_j$ among the orbital degrees of freedom, leading to long-range orbital order. Corresponding microscopic spin-orbital models and their ordered phases have been discussed by Kugel and Khomskii \cite{kk82}.
\end{enumerate}

 We further note that the problem of spinless complex fermions at finite chemical potential coupled to a $\Ztwo$ gauge field, i.e. $\tilde{H}_J^{(2)}+\tilde{H}_h^{(2)}$ in Eq.~\eqref{eq:square_f} was studied in a recent preprint \cite{moroz20}, finding flux-free phases for large fields/chemical potential (however below the critical $h^z \leq 4.0$ for the degenerate phase), consistent with our results.
Moreover the authors argue that the limits $h^z \sim \mu \to \pm \infty$ with zero (one) fermion per site lead to the emergence of a pure even (odd) $\Ztwo$ Ising lattice gauge theory with no (one) $\Ztwo$ background charge per site. We note that this difference emerges \emph{after} projecting to the physical subspace by imposing the local fermion parity constraint $D_j = -1$ on each site.

\subsubsection{$\nu = 3$ model on the honeycomb lattice} \label{sec:pd_hc}

\begin{figure}[!tb]
\includegraphics[width=.9\columnwidth,clip]{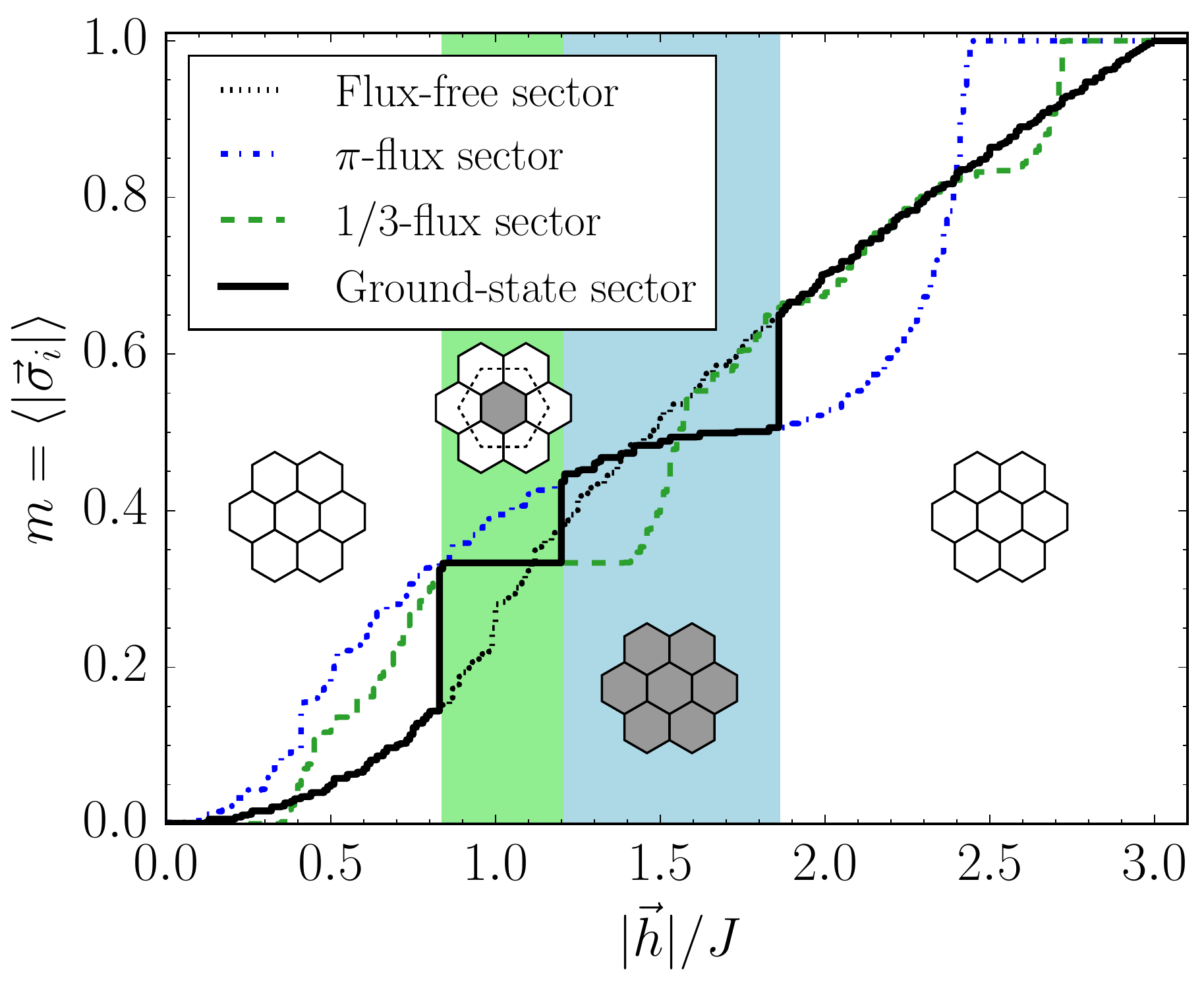}
\caption{
Same as Fig.~\ref{fig:sqMag}, but for the honeycomb spin-orbital liquid as a function of magnetic field $|\vec h|$. 
At low fields, the $0$-flux state is stable (white area). Intermediate fields $0.84 \lesssim |\vec h|/J \lesssim 1.21$ induce a $1/3$-flux crystal (green area) with an enlarged physical unit cell, as indicated by the dashed hexagon, while a $\pi$-flux state is stabilized for $1.21 \lesssim |\vec h|/J \lesssim 1.86$ (blue area). For $|\vec h|/J \gtrsim 1.86$, the ground state is again flux free (white area).}
\label{fig:hcMag}
\end{figure}

We present the ground-state flux configuration and the magnetization curve of the $\nu =3$ honeycomb-lattice model in an external magnetic field in Fig.~\ref{fig:hcMag}.
While the $\nu = 2$ system can be mapped to a tight-binding model of spinless complex fermions, we note that an analogous mapping in the $\nu=3$ model needs to select two out of the three Majorana fermions, such that the model's $\SO(3)$ symmetry is no longer manifest.
For notational clarity, here, we determine the spectrum in a manifestly $\SO(3)$-invariant manner.
We utilize that in crystalline flux sectors we can employ (residual) translational invariance to Fourier-transform the Majorana fermions as
\begin{equation} \label{eq:majorana-ft}
  c_{s,j}^\alpha = \sqrt{\frac{2}{N}} \sum_{\bvec k \in \mathrm{BZ} /2} \left[ {c_{s,\bvec{k}}^{\alpha}} \eu^{\iu \bvec{k} \cdot \bvec{x}_j} + {c_{s,\bvec{k}}^\alpha}^\dagger \eu^{-\iu \bvec{k} \cdot \bvec{x}_j} \right],
\end{equation}
where the Fourier modes ${c_{s,\bvec k}^{\alpha}}$ are canonical fermionic operators in the half Brillouin zone (BZ/2), with flavor index $\alpha = x,y,z$ and sublattice index $s = 1, \dots, N_\mathrm{s}$, where $N_\mathrm{s}$ denotes the number of sites in the Majorana unit cell corresponding to the respective flux sector~%
\footnote{Note that the real-space unit cell of the Majorana problem may be larger than the physical unit cell determined from the flux background due to the projective implementation of translational symmetries, see also Refs.~\cite{kitaev06,batista19}}.
The spectrum of the Hamiltonian can then be found straightforwardly.

In the 0-flux sector, fixing the gauge $u_{ij} = +1$ for all $i \in A$, $j \in B$, the Hamiltonian is then written as
\begin{align}
  \tilde{\mathcal{H}}^{(3)}_J+ \tilde{\mathcal{H}}^{(3)}_h = & - \sum_{\bvec{k} \in \mathrm{BZ}/2}  \psi_\bvec{k}^\dagger \Big\{\big[ \Re f(\bvec{k})\, \Sigma^y \nonumber\\ &+ \Im f(\bvec{k})\, \Sigma^x \big] \otimes \mathds{1}_{3} 
  + \mathds{1}_2 \otimes 2 \vec h \cdot \vec L \Big\} \psi_\bvec{k},
\end{align}
with the six-component spinor $\psi_\bvec k = (c^x_{A,\bvec k}, c^y_{A,\bvec k}, \dots, c^z_{B,\bvec k})^\top$ and $f(\bvec k) = 2 J( 1 + \eu^{\iu \bvec k \cdot \bvec n_1} + \eu^{\iu \bvec k \cdot \bvec n_2})$, where $\bvec n_{1,2} = (\pm \frac{1}{2}, \frac{\sqrt{3}}{2})$ are the honeycomb lattice vectors. $\Sigma^x$ and $\Sigma^y$ denote $2 \times 2$ Pauli matrices.
The spin-1 matrices $\vec L = (L^\alpha)$ have been defined in the context of Eq.~\eqref{eq:majoranaSO3}.
The above Hamiltonian is readily diagonalized, yielding six bands in the half Brillouin zone,
\begin{align} \label{eq:hc-0-flux-spectra}
  \varepsilon_{1,2}(\bvec k) = 2 |\vec h| &\pm |f(\bvec k)|, \quad \varepsilon_{3,4}(\bvec k) = -2 |\vec h| \pm |f(\bvec k)|, \nonumber\\ &\text{and} \quad \varepsilon_{5,6}(\bvec k) = \pm |f(\bvec k)|.
\end{align}
We thus find that a finite magnetic field shifts two Dirac cones of the three dispersing Majoranas away from half filling, leading to Fermi pockets for intermediate field strengths 
\footnote{We can in fact continue $\varepsilon_{3,4}(k)$ to the complementary half of the Brillouin zone so that one may define normal modes on the full Brillouin zone (``unfolding'').
We then obtain two bands of spinless complex fermions on the honeycomb lattice, as one would have obtained by letting $f=(c^x+ \iu c^y)/2$ for fields $\vec h = (0,0,h)$.}%
, while $\varepsilon_{5,6}$ give rise to a single Dirac cone (equivalent to two Majorana cones) that remains at half filling.

In fact, the property of two Majorana bands being independent of the field holds also in other flux sectors:
For general flux configurations, the spinor $\psi_{\bvec{k}}$ has $3N_\mathrm{s}$ components, where $N_\mathrm{s}$ again denotes the number of sites in the corresponding (enlarged) unit cell.
The kinetic energy of the three dispersing Majoranas can then be written in terms of a $3N_\mathrm{s} \times 3N_\mathrm{s}$ matrix $M_{\bvec{k}} \otimes \mathds{1}_{3}$.
Block diagonalizing with some unitary $\mathds{1}_{N_\mathrm{s}} \otimes U$ gives
\begin{align}
    \tilde{\mathcal{H}}^{(3)}_J+ \tilde{\mathcal{H}}^{(3)}_h  & = 
    \sum_{\bvec k \in \mathrm{BZ} / 2} {\psi}_{\bvec{k}}^\dagger \Big[
    \left(M_{\bvec k} + 2 | \vec h| \mathds{1} \right)  \oplus M_{\bvec k}
    \nonumber\\* &\quad
     \oplus \left(M_{\bvec k} - 2 |\vec h| \mathds{1} \right) \Big] {\psi}_{\bvec{k}},
\end{align}
which reveals that $|\vec h|$ acts generally as a chemical potential for two of the three sets of (otherweise identical) fermion bands, while one set of bands remains unaffected by the field in a given flux sector.

As we increase the field strength $|\vec h|$, we find that at $|\vec h| \approx 0.83 J$, there is a first-order transition out of the flux-free sector to a flux-crystal phase with 1/3 flux density, leading to a discontinuity in the magnetization curve, see Fig.~\ref{fig:hcMag}.
We find that in this flux-crystal phase, all Majorana bands become gapped. This is similar to the situation in the $\nu=1$ Kitaev model~\cite{batista19}.
Consequently, as the occupancy of the respective bands does not change for variations in $|\vec h|$, the magnetization $\vec m = \langle\vec \sigma \rangle$ remains constant, leading to a magnetization plateau throughout the $1/3$-flux phase.
Further increasing $|\vec h|$, a first-order transition occurs at $h \approx 1.23 J$, which is associated with the $\pi$-flux state with $W_p = -1$ on all plaquettes $p$ becoming the lowest-energy configuration.
The Majorana dispersion in this sector is gapless, but the magnetization is found to increase only slowly with $|\vec h|$.
Another first-order transition at $|\vec h| \approx 1.88 J$ then gives way to a flux-free ground state at high fields, with an approximately linear increase of the magnetization $|\vec m|$ as a function of $|\vec h|$ up to $|\vec h | = 3 J$, above which the magnetization is fully saturated.
This saturation can be understood by considering the spectra in the flux-free phase given in Eq.~\eqref{eq:hc-0-flux-spectra}: For $|\vec h| > \frac12 \max_{\bvec k} |f(\bvec k)| = 3J$, the Fermi level lies outside the respective bands, and all states associated with two of the three dispersing Majorana flavors can be considered to be fully occupied and empty, respectively.

We emphasize that for $|\vec h| > 3 J$, a gapless band persists with the dispersion $\varepsilon_{5,6}$ in Eq.~\eqref{eq:hc-0-flux-spectra} being independent of $|\vec h|$.
This band leads to the stabilization of the flux-free ground state of the gauge field by Lieb's theorem, and the spin-polarized orbital liquid is protected by the finite flux gap.
This is in contrast to the square-lattice model, which becomes unstable at large fields, as described in the previous subsection.

\subsection{Majorana spectra and topological transitions}

\begin{figure*}
    \includegraphics[width=\textwidth]{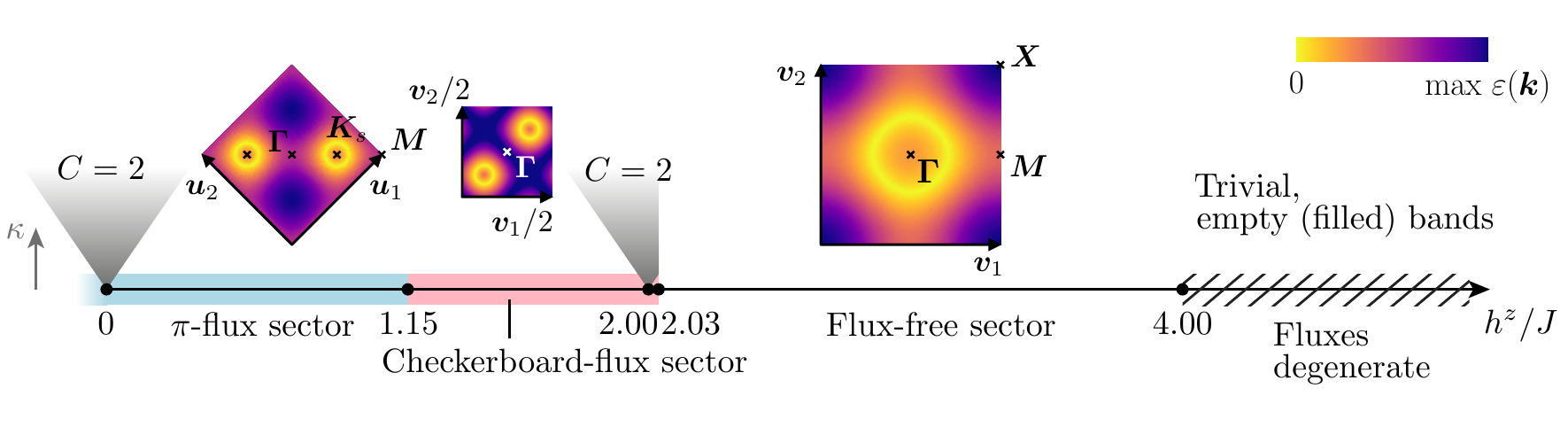}
    \caption{Topological phase diagram of the $\nu = 2$ model on the square lattice as a function of applied external field $h^z$. The grey shaded areas near $h^z = 0$ and $2$ indicate topological phases obtained by applying a small time-reversal-symmetry-breaking perturbation $0<\kappa \ll 1$.
    The spectrum changes discontinuously at first-order transitions between different flux sectors.
    The color plots in the insets show representative dispersions $\varepsilon(\bvec k)$ of the lowest fermion band in the full Brillouin zone for selected values of $h^z/J$ in the Majorana metal phases, using $h^z / J = 0.8$, $1.5$, and $2.5$.
    For $h^z > 4 J$, all quasiparticle bands are empty and all flux sectors become degenerate, which is unstable towards confinement.
    The reciprocal lattice vectors are $\bvec{u}_{1,2} = (\pm \pi,\pi)$ in the $\pi$-flux phase and $\bvec{v}_{1} = (2 \pi,0)$, $\bm{v}_2 = (0, 2 \pi)$ in the $0$-flux phase.}%
   \label{fig:topo_pd_sq}
\end{figure*}

The magnetization curves of the $\nu =2$ and $\nu =3$ models shown in Figs.~\ref{fig:sqMag} and \ref{fig:hcMag} reveal first-order phase transitions between different flux sectors.
While the $1/3$-flux crystal in the honeycomb-lattice model is fully gapped, the other phases have gapless spinon bands.
The gapless phases differ in the momentum-space topology of the respective wave functions in that some of them feature Dirac points that possess a topological charge (``vorticity'').
This implies that, upon adding a small time-reversal-symmetry-breaking perturbation that gaps out the respective Dirac nodes, the system realizes a topologically nontrivial gap and supports chiral edge modes. Such topologically-ordered states realize the sixteen different anyon theories~\cite{CSVJT20}, as classified by Kitaev \cite{kitaev06}.
For the models defined by Eq.~\eqref{eq:h_gamma} at zero external fields, such gaps are opened by including three-site couplings that lead to chiral next-nearest-neighbor hopping, 
\begin{equation} \label{eq:break_TRS}
    \tilde{\mathcal{H}}^{(\nu)}_\kappa = \kappa \sum_{\circlearrowright {\langle ijk \rangle}_{\gamma \gamma'} } u_{ij} u_{jk} \left( \iu c_i c_k
    + \sum_{\beta=\gamma_{\textrm{m}}+1}^{2q+3} i b^\beta_i b^\beta_k \right),
\end{equation}
where ${\circlearrowright{\langle ijk \rangle}_{\gamma \gamma'}}$ refers to clockwise summation over three sites within the same plaquette, where $i$ and $j$
($j$ and $k$) are connected via a $\gamma$-type ($\gamma'$-type) bond~\cite{CSVJT20}.

Following Kitaev \cite{kitaev06}, the gapped topological states can be classified in terms of the Chern number
\begin{equation} \label{eq:chern}
    \Ch = \frac{1}{\pi} \int_{\mathrm{BZ}/2} \du^2 \bvec{k} \tr F_{xy}(\bvec k) \in \mathbb{Z}.
\end{equation}
Here, $F_{xy} \equiv (F^{ab}_{xy})$ denotes the Berry curvature $F^{ab}_{xy}(\bvec k) = \partial_{k_x} A^{ab}_{k_y} - \partial_{k_y} A^{ab}_{k_x} + \iu ([A_x,A_y])^{ab}$ of the non-Abelian Berry connection $\bvec A^{ab} = (A^{ab}_{k_x},A^{ab}_{k_y}) = \langle \psi^a(\bvec k) | (-\iu \nabla_{\bvec{k}}) | \psi^b(\bvec k) \rangle$, where $a,b$ index occupied bands defined in the half Brillouin zone ($\text{BZ}/2$) \cite{zhang04}.
By writing Eq.~\eqref{eq:chern} in terms of the non-Abelian Berry connection, we anticipate that occurring bands are degenerate. Working in the half Brillouin zone allows us to use canonical complex fermionic operators in reciprocal space.
We further note that the above definition for $\Ch$ agrees with Kitaev's convention in the $\nu =1$ case \cite{kitaev06}.
The corresponding canonical Chern number for complex spinless fermions is then given as $\ChD = \frac12 \Ch$. 
Consider, for instance, a single Dirac cone at the $\bvec{K}$ point in $\mathrm{BZ}/2$ (corresponding to two Majorana cones at $\bvec{K}$ and $-\bvec{K}$ in the full Brillouin zone) as in the original $\nu=1$ Kitaev model on the honeycomb lattice. The corresponding topological charge then is  $\Ch = \frac{1}{\pi} \oint_{\mathcal{C}(\bvec K)} \tr \bvec{A} \cdot \du \bvec k \equiv 1 \mod 2$, where we have integrated over a closed path $\mathcal C(\bvec K)$ around $\bvec K$.

For the following discussion, we evaluate Eq.~\eqref{eq:chern} using the Fukui-Hatsugai-Suzuki algorithm \cite{fukui05} for the wave functions obtained by diagonalizing the respective free-fermion problem $\tilde{\mathcal{H}}^{(\nu)}+\tilde{\mathcal{H}}^{(\nu)}_h + \tilde{\mathcal{H}}^{(\nu)}_\kappa $ in momentum space for a given flux sector.
We have also computed the Bott index \cite{Loring_2010} directly for the finite-size systems that were used for finding the lowest-energy flux sector, and we have verified consistency with the momentum-space formulae Eq.~\eqref{eq:chern}.

\subsubsection{$\nu = 2$ model on the square lattice}

On the square lattice, we utilize the mapping to complex spinless fermions, under which the time-reversal-symmetry-breaking perturbation in Eq.~\eqref{eq:break_TRS} becomes
\begin{equation} \label{eq:nnn-square}
    \tilde{\mathcal{H}}^{(\nu)}_\kappa = \kappa \sum_{\circlearrowright {\langle ijk \rangle}_{\gamma \gamma'} } 2 u_{ij} u_{jk} \left( \iu f_i^\dagger f_{k} + \hc \right).
\end{equation}
We can thus analyze the topology of the complex-fermion wavefunctions in the full Brillouin zone, arising from evaluating Eq.~\eqref{eq:square_f} in a given flux background. If a non-trivial gap opens up with finite Chern number $\ChD$, the system's corresponding Majorana Chern number is obtained as $\Ch = 2 \ChD$.
An overview of the topological transitions associated with a change in the free-fermion topology in the respective flux sectors is given in Fig.~\ref{fig:topo_pd_sq}.

At $h=0$, the spectrum of complex fermions posses two Dirac cones at $\bvec K_\mathrm{s} = (\pi/2,0)$ and $\bvec K'_\mathrm{s} =(-\pi/2,0)$ in the full Brillouin zone.
They are gapped out by an small $\kappa>0$, yielding a Majorana Chern number of $\Ch = 2 \ChD = 2 \equiv \nu$ as required by construction for Kitaev's sixteenfold way \cite{kitaev06,CSVJT20}.  
At finite (but small) $h^z > 0$, the Fermi level is shifted away from $f$-fermion particle-hole symmetry, yielding two Fermi pockets encirculating the $\bvec K$ and $\bvec K'$ points and an infinitesimal $0<\kappa \ll 1$ no longer opens up a gap, i.e., the gapless Majorana-metal state is stable.
Increasing $h^z$ further beyond the first-order transition to the checkerboard-flux crystal, the dispersion again features two circular Fermi surfaces encircling $\pm(\pi,\pi)/4$, which shrink to Dirac points upon increasing the field towards $h^z = 2J$.
These Dirac cones at $h^z = 2J$ become gapped out for small finite $\kappa$ and again yield  $C^\mathrm{M} = 2$.
Note that on plaquettes with zero flux, $W_p = +1$, the next-nearest-neighbor hoppings in Eq.~\eqref{eq:nnn-square} interfere destructively, such that for $h^z > 2 J$, when the lowest-energy sector is given by the flux-free ground state and the Fermi level lies inside the cosine-like band, an infinitesimal $\kappa$ does not gap out the dispersion.
For $h^z > 4J$, we find that all bands lie above the Fermi level and thus the fermionic spectrum is trivial, similar to the strong-pairing phase in $p$-wave superconductors~\cite{Read2000}.

\subsubsection{$\nu = 3$ model on the honeycomb lattice}

\begin{figure*}
    \includegraphics[width=\textwidth]{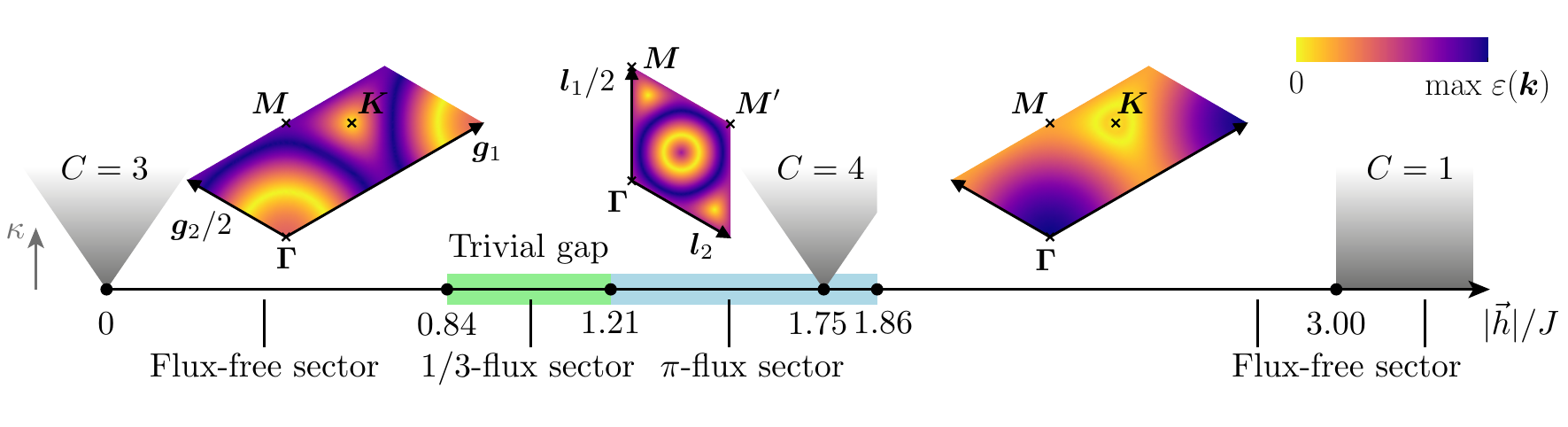}
    \caption{Topological phase diagram of the $\nu = 3$ model on the honeycomb lattice as function of applied external field $|\vec h|$.
    The grey shaded areas near $|\vec h|/J = 0$, $1.75$, and above $3$ indicate topological phases obtained by applying a small time-reversal symmetry-breaking perturbation $0<\kappa\ll 1$.
    The color plots in the insets show representative dispersions $\varepsilon(\bvec k)$ of the lowest fermion band in the half Brillouin zones for selected values of $|\vec h| / J$ in the Majorana metal phases, using $|\vec h|/J = 0.5$, $1.3$, and $2.5$.
   The reciprocal lattice vectors are $\bvec{g}_{1,2} = 2 \pi (\pm 1, \sqrt{3})$ in the $0$-flux phase and $\bvec{l}_1 = (0, 4 \pi /\sqrt{3})$, $\bvec{l}_2 = (\pi,-\pi/\sqrt{3})$ for the $\pi$-flux phase.}%
   \label{fig:topo_pd_hc}
\end{figure*}

The topological phase diagram on the honeycomb lattice is shown in Fig.~\ref{fig:topo_pd_hc}.
At $\vec h=0$, the three degenerate Dirac cones at $\bvec K = (2 \pi/3, 2 \pi/\sqrt{3})$
in $\mathrm{BZ}/2$ become gapped out and give rise to the finite Chern number $\Ch= 3$~\cite{CSVJT20}.
For $0< |\vec h|/J \lesssim 0.84$, two of the three Dirac cones are shifted away from zero energy and thus give rise to a finite Fermi surface, while the third Dirac cone remains protected. Upon breaking time-reversal symmetry with a finite small $\kappa \ll 1$, the Dirac cone is gapped out, while the Fermi-surface state remains gapless.
This is in contrast to the nodal lines found in the $\nu=1$ Kitaev model with additional fourth-nearest neighbor interactions, which become gapped out for infinitesimal $\kappa$ \cite{batista19,halasz20}. 
In the $1/3$-flux crystal, the system possesses a trivial gap, while in the $\pi$-flux crystal for $|\vec h|\gtrsim 1.21 J$, the spectrum in the reduced half Brillouin zone, corresponding to the residual translational symmetry, features two Dirac cones and a Fermi surface.
This Fermi surface is formed by the intersection of a Dirac node at $\bm{M}'/2 = (\pi,\pi/\sqrt{3})/2$, centered at some nonzero elevated energy with its particle-hole-symmetric counterpart.
Upon further increasing the field, these two Dirac cones move to the Fermi level and become degenerate at $|\vec h| = 1.75 J$, at which the Fermi surface shrinks to an isolated point.
Thus, at $|\vec h| = 1.75 J$, an infinitesimal $\kappa$ suffices to gap out the dispersion and gives rise to topological gap with Majorana Chern number $\Ch =4$.
For $1.75 < |\vec h|/J \lesssim 1.86$, the two Dirac cones at $\bm{M}'/2$ again move to higher and lower energies, respectively, and thus a Fermi surface is formed by their intersection, analogous to the case $|\vec h|/J < 1.75$.  
Above $|\vec h|/J \approx 1.86$, the flux-free sector is again stabilized.
As in the low-field limit, the spectrum for $1.86 \lesssim |\vec h|/J < 3$ features a Dirac cone and a Fermi surface that is stable for small $\kappa$.
Upon approaching $|\vec h| = 3 J$, the maxima and minima, respectively, of the two metallic Majorana bands are shifted towards the Fermi level, such that the Fermi surfaces shrink to isolated points with quadratic dispersion at the $\bvec{\Gamma}=(0,0)$ point in BZ/2.
As the perturbation is odd in momentum, this quadratic band touching remains gapless, while the Dirac cone becomes gapped out.
For $h^z > 3 J$, two of the three bands are completely filled and empty, respectively, and only the Dirac cone at $\bm{K}$ remains, yielding $\Ch = 1$ throughout the high-field phase for small $\kappa >0$.

\section{Solvable nearest-neighbor couplings} \label{sec:NN_int}

In this section, we discuss the effect of solvable nearest-neighbor terms as given in Sec.~\ref{sec:nnn-perturb}.
Instead of mapping out the full phase diagram in the respective high-dimensional parameter spaces, we rather highlight characteristic features of individual perturbations.

As a general remark, we note that in particular the inclusion of additional Kitaev- and $\Gamma$-type interactions lead in the limits of large $|K|/J$ and large $|\Gamma|/J$, respectively, to increasingly localized states.
This may be understood by noting that in the Majorana representation, 
such highly-frustrated \cite{baskaran08,kim18} bond-dependent interactions only facilitate disconnected hopping paths for the distinct Majorana flavors, 
$\sigma^\alpha_i \sigma^\beta_j \otimes \dots \mapsto c_i^\alpha c^\beta_j$,
resulting in flat bands in the Majorana spectrum.

\subsection{$\nu = 2$ model on the square lattice}

\subsubsection{Spatially isotropic $\bar\Gamma$ interaction}

We map the spatially isotropic $\bar\Gamma$ interaction in Eq.~\eqref{eq:bargamma-sq} to Majorana fermions, obtaining
\begin{equation}
    \tilde{\mathcal{H}}^{(2)}_J + \tilde{\mathcal{H}}^{(2)}_{\bar\Gamma} = \sum_{\langle ij \rangle} \iu u_{ij} \begin{pmatrix} c^x_i & c^y_i \end{pmatrix} \begin{pmatrix} J & - \bar\Gamma \\ - \bar \Gamma & J \end{pmatrix} \begin{pmatrix} c^x_j \\ c^y_j \end{pmatrix}.
\end{equation}
The above $2 \times 2$ matrix is readily diagonalized by forming new Majorana operators $d^1_i = \left( c^x_i + c^y_i \right)/\sqrt{2}$ and $d^2_i = \left( c^x_i - c^y_i \right)/\sqrt{2}$, yielding the eigenvalues $J\mp\bar{\Gamma}$. It is straightforward to verify that the operators $d^{1,2}_i$ satisfy the Majorana anticommutation relations $\{d^\alpha_i, d^\beta_j \} = 2 \delta^{\alpha \beta}\delta_{ij}$.
The Hamiltonian then maps to a two-flavor Majorana hopping problem on the square lattice in the background of a static $\Ztwo$ gauge field with two different hopping parameters for the two flavors,
\begin{equation} \label{eq:sq-d-h}
    \tilde{\mathcal{H}}^{(2)}_J + \tilde{\mathcal{H}}^{(2)}_{\bar\Gamma} = \sum_{\langle ij \rangle} \iu u_{ij} \left[ \left(J-\bar{\Gamma}\right) d^1_i d^1_j + \left(J+\bar{\Gamma}\right) d^2_i d^2_j \right].
\end{equation}
Using Lieb's theorem for the two individual hopping problems \cite{lieb94}, we deduce that the ground state is always in the $\pi$-flux sector.
Note that, due to the broken $\SO(2) \simeq \Uone$ symmetry, using the mapping to complex fermions introduced in Sec.~\ref{sec:pd_square} on Eq.~\eqref{eq:sq-d-h} would introduce pairing terms $\sim (f^\dagger_i f^\dagger_j + \hc)$.

For $J = \bar\Gamma$ or $J=-\bar\Gamma$, one of the two Majorana modes drops out from Eq.~\eqref{eq:sq-d-h} and thus forms a completely flat band at zero energy. Here, the remaining dispersing Majorana mode still stabilizes the $\pi$-flux ground state. 
The degeneracy of the nondispersing band may either be lifted by further Majorana hopping processes, or makes the system particularly susceptible to spontaneous symmetry breaking upon the inclusion of interactions between the Majoranas~\cite{seifert20}. This interesting direction is left for future work.

\subsection{$\nu = 3$ model on the honeycomb lattice}
\subsubsection{Spatially isotropic $\bar\Gamma$ interaction}

A spatially isotropic flavor-off-diagonal interaction is also possible in the $\nu=3$ model, by assuming $\Gamma = \Gamma' \equiv \bar{\Gamma}$ and $K=0$ in Eq.~\eqref{eq:hkgg}.
The Hamiltonian may then be written in the Majorana representation as
\begin{equation}
    \tilde{\mathcal{H}}^{(3)}_J + \tilde{\mathcal{H}}^{(3)}_{\bar\Gamma} = \sum_{\langle ij \rangle} \iu u_{ij} \bigg[ J \sum_\alpha c^\alpha_i c^\alpha_j - \bar\Gamma \sum_{\alpha < \beta} \left( c^\alpha_i c^\beta_j + c^\beta_i c^\alpha_j \right) \bigg].
\end{equation}
Using arguments similar to those presented in Sec.~\ref{sec:pd_hc}, the Fourier transformation in a given flux sector yields
\begin{equation} \label{eq:iso-gamma}
    \tilde{\mathcal{H}}^{(3)}_J + \tilde{\mathcal{H}}^{(3)}_{\bar\Gamma} = \sum_{\bvec{k} \in \mathrm{BZ}/2} \psi_\bvec{k}^\dagger \Bigg[M_\bvec{k} \otimes \begin{pmatrix} J & -\bar\Gamma & - \bar\Gamma \\ -\bar\Gamma & J & -\bar\Gamma \\ -\bar\Gamma & - \bar\Gamma & J \end{pmatrix}   \Bigg] \psi_\bvec{k},
\end{equation}
where $\psi_\bvec{k} = (c_{A,\bvec k}^x, c_{A, \bvec k}^y, \dots, c_{N_\mathrm{s},\bvec k}^z)^\top$.
By means of a global unitary transformation $\psi_\bvec{k} \mapsto (\mathds{1}_{N_\mathrm{s}} \otimes U) \psi_\bvec{k}$ with the unitary $3 \times 3$ matrix $U$, the square bracket $[\,\cdots]$ on the right-hand-side of Eq.~\eqref{eq:iso-gamma} can be block-diagonalized,
\begin{equation} \label{eq:iso-gamma-blockdiag}
    [\,\cdots] = (J-2 \bar\Gamma) M_\bvec{k} \oplus (J + \bar\Gamma) M_\bvec{k} \oplus (J+ \bar\Gamma) M_\bvec{k}. 
\end{equation}
The dispersion hence decomposes into three blocks with identical momentum dependence, but different prefactors.
As each block is proportional to the Hamiltonian of free Majorana fermions hopping on the honeycomb lattice in the background of the respective flux configuration, it follows that the ground state of Eq.~\eqref{eq:iso-gamma} is in the flux-free sector and thus $M_\bvec k = - \Re f(\bvec k) \Sigma^y - \Im f(\bvec k) \Sigma^x$, as in Sec.~\ref{sec:pd_hc} for vanishing fields.

Equation~\eqref{eq:iso-gamma-blockdiag} reveals that at $J = -\bar\Gamma$ ($J = 2 \bar\Gamma$) two blocks (one block) vanish(es) identically and thus give(s) rise to four (two) degenerate flat bands at zero energy. The remaining bands realize one (two) Dirac node(s), in analogy to the result on the square-lattice model.

For all other values of $\Gamma/J$, the dispersion features three Dirac cones, two of which have an identical Fermi velocity due to two Majorana flavors being degenerate. As can be seen from Eq.~\eqref{eq:iso-gamma-blockdiag}, $\tilde{\mathcal{H}}^{(3)}_J + \tilde{\mathcal{H}}_{\bar\Gamma}^{(3)}$ has a hidden $\mathrm{O}(2)$ symmetry that mixes the two degenerate modes.

\subsubsection{Bond-dependent diagonal $K$ interaction} \label{sec:hc_k}

\begin{figure}[!tb]
\includegraphics[width=.9\columnwidth,clip]{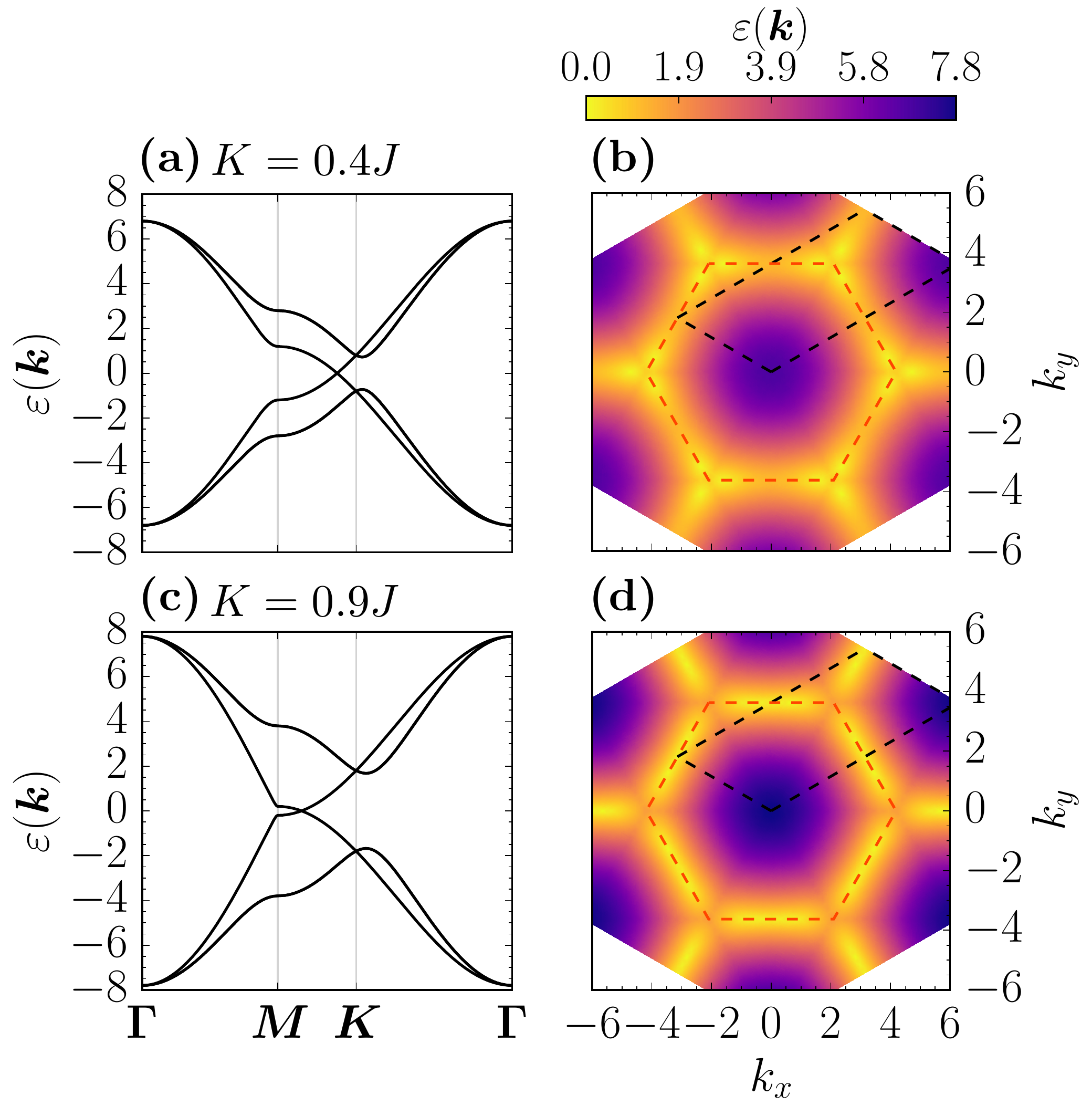}
\caption{
(a) Majorana dispersion along high-symmetry path in the Brillouin zone for the $J$-$K$ model on the honeycomb lattice defined by $\tilde{\mathcal{H}}^{(3)}_J + \tilde{\mathcal{H}}^{(3)}_K$ for $K = 0.4 J$.
(b) Dispersion of the lowest-energy band for $K = 0.4 J$.
Red dashed lines denote the hexagonal Brillouin zone, black dashed lines denote $\mathrm{BZ}/2$ as used in Fig.~\ref{fig:topo_pd_hc}.
(c) Same as (a) for $K=0.9J$, showing that the Dirac cones move away from the $\bvec K$ point towards the $\bvec M$ points. (d) Same as (b) for $K=0.9J$.}
\label{fig:disp_kit_HC}
\end{figure}

The $\nu = 3$ model allows additional spin-orbital interactions that are bond dependent also in the spin sector and preserve solvability. Taking $\Gamma = \Gamma' = 0$ in Eq.~\eqref{eq:hkgg} and transforming to the Majorana representation yields
\begin{align}
    \tilde{\mathcal{H}}^{(3)}_J + \tilde{\mathcal{H}}^{(3)}_K &= 
    \sum_{\langle ij \rangle_\gamma} \iu u_{ij}  \left[(J+ K) c^\gamma_{i} c^\gamma_j + \sum_{\alpha\neq \gamma} J c^\alpha_i c^\alpha_j \right]. \label{eq:j-k-maj}
\end{align}
Clearly, a finite $K$ spoils the system's $\SO(3)$ spin-rotational symmetry.
From the Majorana representation, it is apparent that in the limit $K /J \to \infty$ the $\gamma$-type Majorana fermions are localized at the respective $\gamma$-type bonds, leading to a gapped dispersion and flat bands.
For intermediate values of $K/J$, the ground state remains in the flux-free sector. Using Eq.~\eqref{eq:majorana-ft}, the spectrum is found to be given by $\varepsilon_\pm^\alpha(\bvec k) = \pm |f(\bvec k) +2 K \eu^{\iu \bvec k \cdot \bvec \delta_\alpha}|$, with $\alpha = 1,2,3$ and $(\bvec \delta_1, \bvec \delta_2, \bvec \delta_3)=(\bvec n_1,\bvec n_2, \bvec 0)$. The spin-orbital interaction parametrized by $K$ is thus a $C_3$-symmetric version of the bond anisotropy discussed by Kitaev \cite{kitaev06}.
For small $K/J$, each of the three Dirac cones moves away from the $\bvec{K}$ point along three inequivalent directions in momentum space, as shown in Fig.~\ref{fig:disp_kit_HC}. For $K/J = 1$, they gap out at the $\bvec{M}$ points by merging with the Dirac cones from the other half of the full Brillouin zone.
Generalizing Kitaev's arguments \cite{kitaev06}, the system at $K \gg J$ may be mapped to the toric code \cite{kitaev03}, which is described by an exactly soluble $\Ztwo$ gauge theory with a full gap.

\subsubsection{Bond-dependent off-diagonal $\Gamma$ interaction} \label{sec:hc_gamma}

\begin{figure}[!tb]
\includegraphics[width=.9\columnwidth,clip]{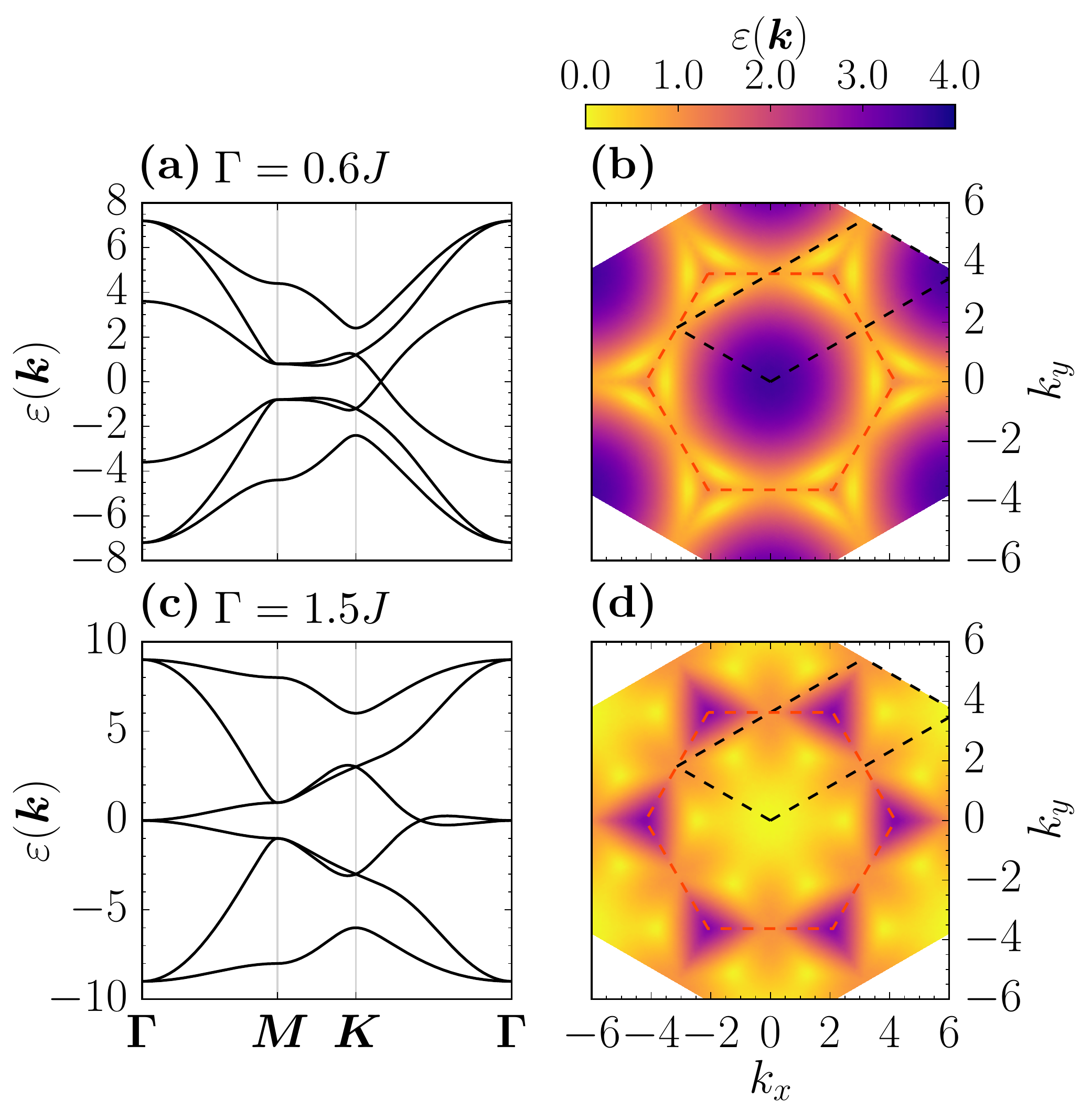}
\caption{
(a) Majorana dispersion along high-symmetry path in the Brillouin zone for the $J$-$\Gamma$ model on the honeycomb lattice defined by $\mathcal{H}^{(3)}_J + \mathcal{H}^{(3)}_\Gamma$, using $\Gamma/J = 0.6$.
(b) Dispersion of the lowest-energy band for $\Gamma/J = 0.6$. Red dashed lines denote the hexagonal Brillouin zone, black dashed lines denote $\mathrm{BZ}/2$ as used in Fig.~\ref{fig:topo_pd_hc}.
(c) Same as (a) for $\Gamma/J = 1.5$.
(d) Same as (b) for $\Gamma/J = 1.5$.
}
\label{fig:disp_gamma_HC}
\end{figure}

Considering $\Gamma' = K \equiv 0$ in Eq.~\eqref{eq:hkgg}, the remaining perturbation to the $\SO(3)$-symmetric model on the honeycomb lattice is given by off-diagonal exchange interactions. In the Majorana representation, the perturbed Hamiltonian reads
\begin{align}
    \tilde{\mathcal{H}}^{(3)}_J + \tilde{\mathcal{H}}^{(3)}_\Gamma = 
    \sum_{\langle ij \rangle_{\gamma}} \iu u_{ij} \left[ J c^\gamma_i c^\gamma_j - \Gamma (c^\alpha_i c^\beta_j + c^\beta_i c^\alpha_j) \right], \label{eq:j-gamma}
\end{align}
where $(\alpha,\beta,\gamma) = (y,z,x)$, $(z,x,y)$, and $(x,y,z)$ on $x$, $y$, and $z$ bonds, respectively.
Performing a variational study for various values of $\Gamma/J$, we find that the $0$-flux sector has the lowest ground-state energy, except for a small parameter window centered at $\Gamma / J = 1$, where the lowest flux sector is given by a ``stripy'' flux pattern with $1/4$ flux density, 
cf.\ Fig.~\ref{fig:fluxconfigs_hc}.
We note, however, that several other flux sectors, including the $1/4$-flux crystal and the flux-free phase, are close in energy at this point, with relative differences $\Delta E / E \sim 10^{-6}$ (see also Fig.~\ref{fig:gs_energs_gamma} in the Appendix for more details on the flux sectors' energies),
requiring a systematic study with higher numerical accuracy to fully resolve the ground state near $\Gamma/J \simeq 1$. This is left for future work.
Here, we instead describe the features of the Majorana dispersion in the $0$-flux sector as a function of $\Gamma/J$:
For small $\Gamma/J$, we find that the Dirac cones move away from the $\bvec{K}$ points towards the center $\bvec\Gamma$ of the Brillouin zone, as shown in Fig.~\ref{fig:disp_gamma_HC}(a,b).
For the particular value of $\Gamma / J = 1$, the flux-free sector features nodal lines. This large number of gapless degrees of freedom is energetically unfavourable and explains the fact that several other flux sectors are found to have competing energies near this point.
At $\Gamma/J = 1.5$, the dispersion features an additional quadratic band touching at the $\bvec\Gamma$ point, see Fig.~\ref{fig:disp_gamma_HC}(c,d).
Increasing $\Gamma/J$, the quadratic band touching at $\Gamma$ splits into Dirac cones, which move along the high-symmetry lines $\bm{\Gamma}$-$\bm{K}$ and $\bm{\Gamma}$-$\bm{K}'$.
At $\Gamma/J = 1.60$, these Dirac points annihilate with the Dirac cones that have moved from the $\bm{K}$ and $\bm{K}'$ points, and the Majorana dispersion becomes fully gapped.
Similar to the case discussed in the previous subsection, we expect that in this limit, integrating out the gapped itinerant fermions yields an Abelian $\Ztwo$ gauge theory.
In the limit $\Gamma/J \to \infty$, flat bands are formed, corresponding to the localization of the itinerant Majorana fermions, because the $\Gamma$-interaction does not facilitate hopping of $\gamma$-type Majorana fermions along a $\gamma$-type bond.

\section{Generalization to $\mathrm{SO}(\nu)$ models}\label{sec:generalization}

In this section, we generalize some of the above results to the $\SO(\nu)$-symmetric models with arbitrary $\nu > 1$ \cite{CSVJT20}. (For a detailed discussion of the $\nu=2$ and $\nu=3$ models we refer the reader to Secs.~\ref{sec:field} and \ref{sec:NN_int}.)
We again start by discussing solvable onsite terms, which can be understood as generalized Zeeman couplings to an external field.

\subsection{Onsite perturbations}

Onsite terms that couple the itinerant Majorana fermions in Eq.~\eqref{eq:h_gamma_u} to external fields can be written as linear combinations of $\sum_j \Gamma^\alpha_j$ and $\sum_j \Gamma^{\alpha \beta}_j$ with $\alpha,\beta = \gamma_\mathrm{m}+1, \dots, 2q+3$, $\alpha<\beta$, where $\gamma_\mathrm{m}=4$ $(3)$ in the square-lattice (honeycomb-lattice) model for even $\nu = 2q$ (odd $\nu=2q+1$). These are the generators of the $\SO(\nu)$ symmetry.
Mapping to Majorana fermions and relabelling the itinerant Majorana fermions in analogy to Sec.~\ref{sec:modelsymm}, the onsite terms map onto
\begin{equation} \label{eq:onsite_nu}
    \tilde{\mathcal{H}}^{(\nu)}_h = - \sum_j \sum_{a < b} h^{ab} \frac{1}{2} {\underline{c}}^\top_j L^{ab} {\underline{c}}_j
\end{equation}
with real coupling constants $h^{ab}$, $1 \leq a < b \leq \nu$, which can be understood as generalized Zeeman field strengths. Here, the $L^{ab}$ are $\nu(\nu-1)/2$ traceless antisymmetric $\nu \times \nu$ matrices, which form a $\SO(\nu)$ algebra in the defining representation, and ${\underline{c}}_j \equiv (c^1_j, \dots, c^\nu_j)^\top$ denotes a $\nu$-dimensional Majorana spinor.
However, since the different $L^{ab}$ do not commute in general, distinct choices of $h^{ab}$ may be related by $\SO(\nu)$ symmetry.
We may therefore, without loss of generality \cite{brauner11}, focus on the maximal set of the commuting generators $[H^p,H^{p'}]=0$ spanned by the Cartan subalgebra $\{H^p\}_{p=1,\dots,q}$ of $\SO(\nu)$.

For even $\nu = 2q$, the Cartan generators are $2q \times 2q$ matrices consisting of $q$ blocks with the Pauli matrix $\Sigma^y$ inserted in the $p$-th block and otherwise zero matrices,
\begin{equation}
    H^p = 0^{2 \times 2}_1 \oplus \dots \oplus \Sigma^y_p \oplus \dots \oplus 0^{2 \times 2}_{q}.
\end{equation}
For odd $\nu = 2q+1$, the Cartan generators are $(2q+1) \times (2q+1)$ matrices that contain an additional $1$-dimensional block,
\begin{equation} \label{eq:cartan_2q+1}
    H^p = 0^{2 \times 2}_1 \oplus \dots \oplus \Sigma^y_p \oplus \dots \oplus 0^{2 \times 2}_{q} \oplus 0^{1\times1}_{q+1}.
\end{equation}
Hence, the onsite Hamiltonian in Eq.~\eqref{eq:onsite_nu} is uniquely written as
\begin{equation}
    \tilde{\mathcal{H}}^{(\nu)}_h = -\sum_j \sum_{p=1}^q h^p \frac{1}{2} {\underline{c}}_j^\top H^p {\underline{c}}_j,
\end{equation}
where the $h^p$, $p=1,\dots,q$, are now field strengths associated with the Cartan generators and can be expressed in terms of the $h^{ab}$ introduced above.
Crucially, each Cartan generator contains exactly one block with $\Sigma^y_p$. This suggests to pair the $(2p-1)$-th and $2p$-th Majorana flavors at each site into a complex fermion $f^{p}_j = (c^{2p-1}_j + \iu c^{2p}_j) /2$, yielding
\begin{equation}
    \frac{1}{2} {\underline{c}}^\top_j \left[ 0^{2 \times 2}_1 \oplus \dots \oplus \Sigma^y_p \oplus \dots \right] {\underline{c}}_j = - \left( 2 {f^p_j}^\dagger f^p_j - 1 \right).
\end{equation}

\subsubsection{$\nu = 2q$ models on the square lattice}

For the $\nu = 2q$ model, the onsite terms together with the unperturbed Hamiltonian in Eq.~\eqref{eq:h_gamma_u} map onto $q$ bands of complex fermions coupled to the $\Ztwo$ background gauge fields,
\begin{align}
    \tilde{\mathcal{H}}^{(2q)}+\tilde{\mathcal{H}}^{(2q)}_h = \sum_{p=1}^q \bigg[ J &\sum_{\langle ij \rangle} u_{ij} \Big( 2 \iu  {f^p_i}^\dagger f^p_j + \hc \Big) \nonumber\\ &+ h^p \sum_i \Big(2 {f^p_i}^\dagger f^p_i -1 \Big) \bigg],
\end{align}
with the generalized field strengths $h^p$ corresponding to a band-dependent chemical potential.
Choosing one particular $h^{r} \geq 4J$ and $h^p=0$ for all $p \neq r$ and using the results of Sec.~\ref{sec:pd_square} leads to all $f^r$-fermion states being unoccupied. The system thus realizes a generalized Kitaev model \cite{CSVJT20} with $(\nu-2)$ itinerant Majorana fermions and a residual $\SO(\nu-2) \times \SO(2)$ symmetry. Note that for each finite $h^p$ there is a residual $\SO(2) \simeq \Uone$ symmetry generated by $H^p$, which cannot be broken explicitly by solvable onsite terms. (This may be achieved, for instance, by choosing distinct Kitaev couplings for the different Majorana flavors, which generates nearest-neighbor pairing terms $\sim f^p_i f^p_j$.)
Choosing $h^p \geq 4 J$ for all $p=1,\dots,q$, all bands are shifted above the Fermi level, and the $\Ztwo$ gauge field is unstable towards confinement due to the degeneracy of all flux configurations, unless explicitly stabilized by additional interactions that preserve the gauge structure.
This general result is consistent with our findings for the $\nu=2$ model.

\subsubsection{$\nu = 2q+1$ models on the honeycomb lattice}

For the $\nu = 2q+1$ model, the above mapping to spinless complex fermions can be performed analogously for the first $q$ blocks of the Cartan generators by pairing the first $2q$ entries of the $(2q+1)$-dimensional Majorana spinor. However, there remains a single Majorana fermion $c^{2p+1}$ associated with the zero weight state of $\SO(2q+1)$, such that the Hamiltonian reads
\begin{align}
    \tilde{\mathcal{H}}^{(2q+1)}&+\tilde{\mathcal{H}}^{(2q+1)}_h = \sum_{p=1}^q \bigg[ J \sum_{\langle ij \rangle} u_{ij} \Big( 2 \iu  {f^p_i}^\dagger f^p_j + \hc \Big) \nonumber\\
    &+ h^p \sum_i \Big(2 {f^p_i}^\dagger f^p_i -1 \Big) \bigg] + J \sum_{\langle ij\rangle} \iu u_{ij} c^{2p+1}_i c^{2p+1}_j. \label{eq:h2q+1_fields}
\end{align}
Choosing any one of the fields $h^r \geq 3J$ and using the results from Sec.~\ref{sec:pd_hc} leads to the associated $f^r$ bands becoming fully unoccupied. This way, one again obtains a generalized Kitaev model with $(\nu-2)$ itinerant Majorana fermions, with the symmetry group reduced as $\SO(\nu) \to \SO(\nu-2) \times \SO(2)$, in analogy to the square-lattice case.
Importantly, fully breaking the $\SO(\nu)$ symmetry by choosing $h^p \geq 3 J$ for $p = 1, \dots, q$ moves all complex modes above the Fermi level, but leaves a single Majorana fermion at zero energy. The latter corresponds to the last term in Eq.~\eqref{eq:h2q+1_fields}, which is invariant under variations of the $h^p$, giving way to a $\nu = 1$ Kitaev orbital liquid ground state in this phase.
Thus, the flux-free ground state is stabilized for sufficiently strong fields $h^p \gg J$ for all $p$, in agreement with our findings in the $\nu=3$ model.

\subsection{Nearest-neighbor perturbations} \label{sec:general-gamma}

We now discuss the effects of solvable nearest-neighbor interactions which break the global $\SO(\nu)$ rotation symmetry in the generalized models.
For simplicity, we focus only the generalization of the $\bar\Gamma$ interaction which can be realized both in the square-lattice as well as honeycomb-lattice models.
We expect that our analysis can be readily extended for other $\SO(\nu)$-breaking perturbations which can be defined for arbitrary $\nu$.

The effects of the generalized $\bar\Gamma$ interaction for the cases of the $\nu=2$ and $\nu=3$ models can be understood within a group-theoretical analysis:
To this end, we note that the presence of a finite $\bar\Gamma$ breaks the global $\mathrm{O}(2)$ symmetry in the $\nu=2$ model on the square lattice.
However, the symmetric group of two elements $\mathcal{S}_2 \simeq \mathbb{Z}_2 \subset \mathrm{O}(2)$ remains a symmetry, which acts in a two-dimensional reducible representation on $(c^x,c^y)^\top$.
Similarly, in the $\nu=3$ model on the honeycomb lattice, we analogously find that the remaining $\mathcal{S}_3 \subset \mathrm{O}(3)$ symmetry acts in the three-dimensional reducible representation on the flavor degrees of freedom $(c^x,c^y,c^z)^\top$. 
\mbox{(Block-)}diagonalizing the $\nu=2$ ($\nu=3$) Hamiltonian in Eq.~\eqref{eq:sq-d-h} [Eq.~\eqref{eq:iso-gamma-blockdiag}] is then equivalent to splitting the reducible representation into the trivial one-dimensional irreducible representation and a further one-dimensional (two-dimensional) irreducible representation.
The trivial representation has eigenvector $(1,1)^\top/\sqrt{2}$ [$(1,1,1)^\top/\sqrt{3}$], while the additional one-dimensional (two-dimensional) irreducible representation is $(1,-1)^\top/\sqrt{2}$ [spanned by $\{(2,-1,-1)^\top/\sqrt{6},(0,1,-1)^\top/\sqrt{2}\}$, corresponding to the twofold degenerate Majorana modes].

The above group-theoretical understanding allows the generalization of the results to arbitrary $\nu$.
The generalized $\bar\Gamma$ interaction is given by
\begin{align}
    \mathcal{H}^{(\nu)}_{\bar\Gamma} = \bar\Gamma \sum_{\langle ij \rangle_\gamma} &\sum_{\alpha=\gamma_\mathrm{m}+1}^{2q+3} \Bigg[ \Gamma^\gamma_i \Gamma^{\gamma\alpha}_j + \Gamma^{\gamma\alpha}_i \Gamma^{\gamma}_j \nonumber\\ &+ \sum_{\beta < \alpha} \left( \Gamma^{\gamma\alpha}_i \Gamma^{\gamma\beta}_j + \Gamma^{\gamma\beta}_i \Gamma^{\gamma\alpha}_j \right) \Bigg].
\end{align}
where again $\gamma_\mathrm{m} = 4$ $(3)$ in the square-lattice (honeycomb-lattice) model for even $\nu=2q$ (odd $\nu=2q+1$).
This term breaks the $\mathrm{O}(\nu)$ symmetry, while the symmetric group $\mathcal{S}_\nu$ of $\nu$ elements remains a symmetry that acts in the  $\nu$-dimensional reducible \emph{natural} representation on the itinerant Majorana fermions.

Mapping to Majorana fermions and introducing a $\nu$-component spinor, it becomes clear that the task of block-diagonalizing the hopping Hamiltonian defined by $\tilde{\mathcal{H}}^{(\nu)}_{J} + \tilde{\mathcal{H}}^{(\nu)}_{\bar\Gamma}$ requires diagonalizing the $\nu \times \nu$ matrix $F = (F_{\alpha\beta})$ with components $F_{\alpha\beta} = J \delta_{\alpha\beta} + (\delta_{\alpha\beta}-1) \bar{\Gamma}$, $\alpha,\beta=1,\dots, \nu$.
Using the generalized matrix determinant lemma, it is easy to see that $F$ possesses the eigenvalues
\begin{equation}
    \lambda_1 = J - (\nu-1) \bar\Gamma \quad \text{and} \quad \lambda_i =  J + \bar\Gamma \quad i=2,\dots,\nu.
\end{equation}
The corresponding one-dimensional eigenspace given by $(1, \dots, 1)$ is the \emph{trivial} representation, while the orthogonal complement, associated with the eigenvalues $\lambda_{2,\dots,\nu}$, defines the $(\nu-1)$-dimensional irreducible \emph{standard} representation of $\mathcal{S}_\nu$. We note that there is an $O(\nu-1)$ symmetry corresponding to global basis rotation in the $(\nu-1)$-dimensional degenerate subspace. 

We thus conclude that $\tilde{\mathcal{H}}^{(\nu)}_{J} + \tilde{\mathcal{H}}^{(\nu)}_{\bar\Gamma}$ features, for arbitrary values of $\bar{\Gamma} / J$, $\nu$ Dirac cones, out of which $(\nu-1)$ are degenerate.
For the special case of $J = (\nu-1) \bar\Gamma$, a single zero-energy flat band and $(\nu-1)$ degenerate Dirac cones are formed, while for $\bar\Gamma = -J$ the spectrum consists of a single Dirac cone and a $(\nu-1)$-fold-degenerate flat band at zero energy.
Specifying to $\nu=2$ and $\nu=3$, respectively, these general results agree with our previous explicit findings for the perturbed $\SO(2)$- and $\SO(3)$-symmetric models.

\section{Discussion and outlook} \label{sec: conc}

In this work, we have considered spin-orbital models on the square and honeycomb lattices that exhibit quantum-spin-orbital-liquid ground states.
They can be solved exactly in analogy to Kitaev's honeycomb model~\cite{kitaev06} and feature Majorana fermions hopping in the background of a static $\Ztwo$ gauge field.
Guided by gauge invariance and symmetry considerations, we have systematically investigated the possible onsite and nearest-neighbor interaction terms that preserve the solvability of the models.
In particular, we have studied in detail the physically important case of external magnetic fields that couple to the spin degrees of freedom.
These induce a series of metamagnetic transitions and stabilize Majorana-Fermi-surface states, as well as semimetallic states with Dirac and/or quadratic band touching points in the Majorana spectrum.
Upon applying small time-reversal-symmetry-breaking perturbations, one can open up a topologically nontrivial band gap for the Majorana fermions. The resulting low-energy theories are classified in Kitaev's sixteenfold way \cite{kitaev06, CSVJT20}.
In the limit of strong magnetic fields, the spin-orbital models are field-polarized in the spin sector, while the orbital degrees of freedom are either in a macroscopically degenerate state on the square lattice or realize a single-Majorana Kitaev orbital liquid on the honeycomb lattice.

\subsection{Non-solvable perturbations} \label{subsec: non-sol}

While materials harboring two-dimensional Kitaev spin-orbital liquids have so far not been uniquely identified, three-dimensional double perovskites such as $\mathrm{Ba}_2\mathrm{YMoO}_6$ realize similar bond-dependent interactions \cite{balents10,deVries10,natori16}.
We emphasize that in candidate materials, a sizeable spin-orbit coupling will be present, such that an external field invariably also couples to the orbital degrees of freedom. This leads to additional onsite terms of the form $\sim \sigma^\alpha_i \otimes \tau^\beta_i$, 
as also obtained from symmetry arguments for $j_\mathrm{eff}=3/2$ systems \cite{natori20}. 
Moreover, the presence of further inter-orbital interactions $\sim \tau^\alpha_i \tau^\beta_j$ may be generically expected in spin-orbital systems \cite{kk82}.
As discussed by Kugel and Khomskii, external pressure may give further rise to terms involving a single orbital operator and could thus ``polarize'' the orbitals \cite{kk82}. 
All those operators generally do not commute with the plaquette flux operators $W_p$ and hence lead to dynamics of the $\Ztwo$ gauge field resulting e.g.~in the disperson of visons. This is also visible from the Majorana representation [see e.g.~Eq.~\eqref{eq:tauOps}], which places $b^{\alpha}$ Majoranas on non-$\alpha$ bonds, rendering a rewriting in terms of local $u_{ij}$ gauge field variables impossible.
These terms are expected to eventually drive confinement transition which are interesting subjects for further study and received recent attention \cite{moroz20,Gazit18,takahashi21}.
Note that there are perturbations that \emph{do} commute with the plaquette operators (such as $\mathds{1} \otimes \tau^{\alpha}\tau^{\alpha}$ on $\langle ij \rangle = \alpha$-links of the honeycomb lattice), but are \emph{not exactly solvable}: From the Majorana representation \eqref{eq:tauOps} it becomes apparent that they either give rise to terms which involve $b^{\alpha}$ Majoranas on $\langle ij \rangle \neq \alpha$ bonds (and thus cannot be rewritten in terms of the gauge field $u_{ij}$), or lead to quartic (or higher) fermion interactions -- an example of the latter case has recently been studied in Ref.~\cite{seifert20}.
If these non-solvable perturbations are small compared to $J$, they can in principle be analysed in perturbation theory and projected to the ground-state flux sector.
We expect the most relevant (by power-counting) terms to modify the dispersion of the itinerant Majorana fermions and lead to hybridization of the respective flavors, in similarity to Kitaev's analysis of an applied magnetic field in the $S=1/2$ honeycomb model \cite{kitaev06}.
Furthermore we note that the only gauge-invariant terms to be generated which \emph{do not} involve the itinerant Majorana fermions will be given by Wilson loop operators of the gauge field $u_{ij}$.
The study of these more realistic models beyond the perturbative regime requires appropriate numerical techniques.
In this sense, our study reveals general features of spin-orbital liquids in appropriately fine-tuned models that preserve exact solvability, and as such can serve as useful starting points for more detailed investigations of material-specific models. 

\subsection{Outlook} \label{subsec: outlook}

Our study opens up several avenues for further theoretical work:
The models constructed host a variety of different parton states, including Majorana metals, semimetallic states, or fully gapped states.
These should be expected to yield distinct features in the corresponding thermalization processes, which can be mapped out using sign-problem-free quantum Monte Carlo methods in the Majorana basis \cite{motome15,eschmann2020}:
For example, while the original Kitaev model shows a characteristic two-crossover behavior that is indicative of the fractionalization process \cite{do17}, we expect that in the $\nu =3$ model with strong magnetic fields $|\vec h| \gg J$, a third peak in the specific heat $C_v$ will occur at temperatures on the order of $|\vec h|$, associated with a thermal disordering of the field-polarized spins.
Furthermore, in this regime, the lowest-temperature crossover associated with flux ordering will take place at lower temperatures as compared to the $\vec h = 0$ case, since the flux gap is reduced by a factor of $1/3$.
We note that a theoretical work in a similar spirit was recently carried out in Ref.~\cite{eschmann2020}, in which the finite-temperature behavior of a $j_\mathrm{eff}=3/2$ Kitaev-type spin-orbital liquid on the Shastry-Sutherland lattice was studied.

Moreover, the fact that Zeeman fields, as solvable onsite perturbations, lead to Majorana-metal ground states in a wide parameter regime allows for intriguing further directions:
$\Ztwo$ spin liquids with Majorana Fermi surfaces have been previously found as ground states of the spin-$1/2$ Kitaev model on two-dimensional lattices in the presence of additional perturbations \cite{Zhang2019,chari20}, as well as on three-dimensional lattices \cite{trebst14,rosch15}.
In the latter case, it was found that these Majorana Fermi surfaces are generically unstable upon the inclusion of interactions, breaking some of the spatial symmetries of the system and giving way to line nodes.
A study of the possible instabilities of the Majorana-Fermi-surface states and the resulting symmetry-broken phases in our two-dimensional spin-orbital models, augmented by generic interactions, appears to be similarly promising.

\textit{Note:} After the completion of this work Ref.~\cite{moroz20} appeared in which the phases of spinless fermions coupled to $\Ztwo$ lattice gauge theory were studied. The model reduces (in the limit of a static gauge field) to the (exact) parton construction of our square-lattice quantum spin-orbital liquid, yielding results largely consisted with our findings.

\acknowledgements

We thank Carsten Timm, Frank Schindler and Sergej Moroz for helpful discussions. This research has been supported by the Deutsche Forschungsgemeinschaft (DFG) through SFB 1143 (project id 247310070) and the W\"{u}rzburg-Dresden Cluster of Excellence {\it ct.qmat} (EXC 2147, project id 390858490). S.C.\ acknowledges funding by the IMPRS for Many Particle Systems in Structured Environment at MPI-PKS. The work of L.J.\ is funded by the DFG through the Emmy Noether program (JA2306/4-1, project id 411750675).

\appendix

\section{Flux configurations used in variational treatment}

\begin{figure*}[!tb]
    \includegraphics[width=.90\textwidth]{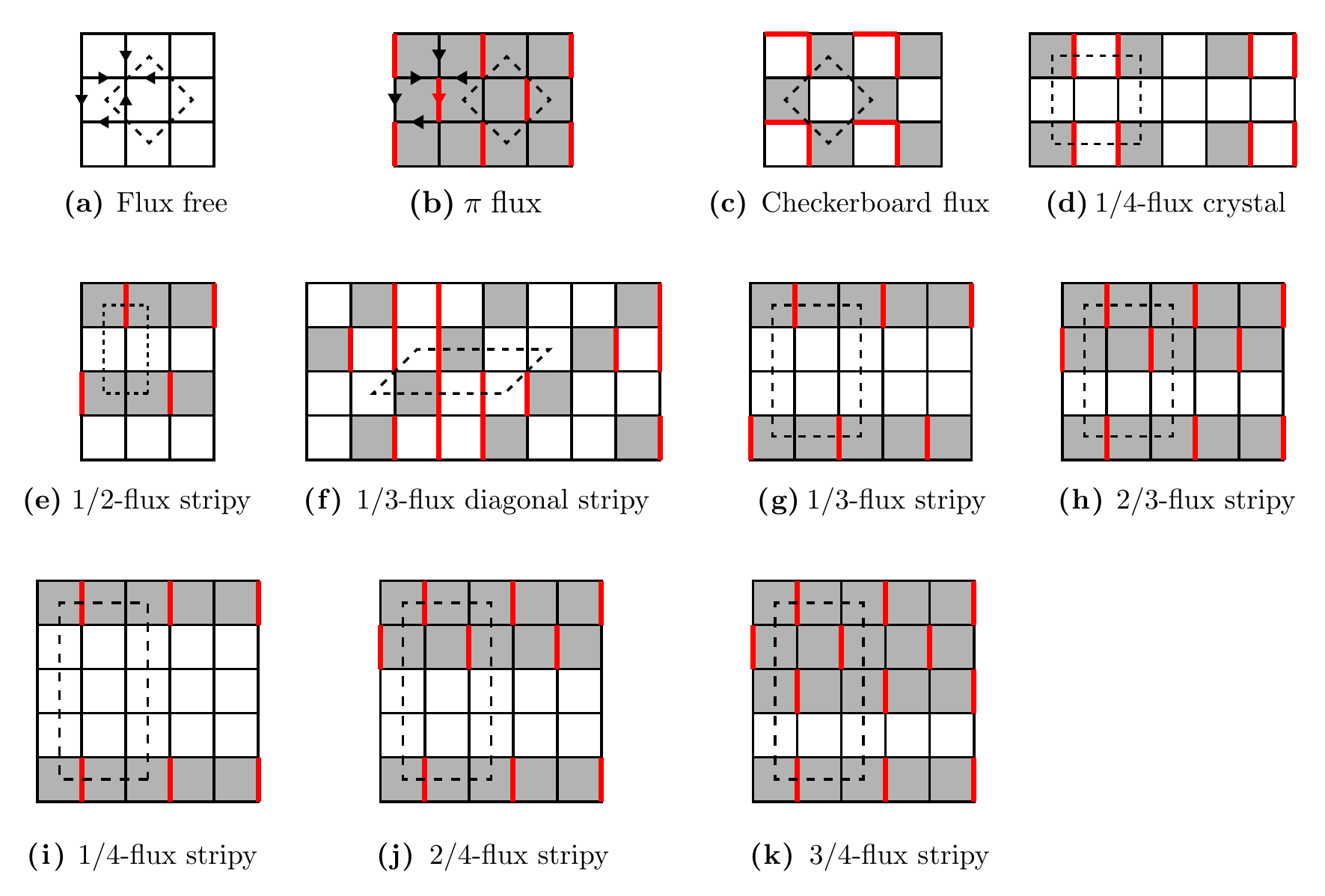}
    \caption{Flux patterns considered in the variational determination of the ground state for the $\nu=2$ model, and a representative fixed configuration of the gauge field $u_{ij}$.
    Red (bold) bonds indicate ``flipped'' bonds $u_{ij} = -1$ on the background of the canonical flux-free gauge configuration $u_{ij}=+1$ for $i \in A$, $j \in B$ sublattices, as indicated by the arrows in panels (a) and (b).
    Dashed lines denote the physical unit cells.
    }
   \label{fig:fluxconfigs_sq}
\end{figure*}

\begin{figure*}[!tb]
    \includegraphics[width=.90\textwidth]{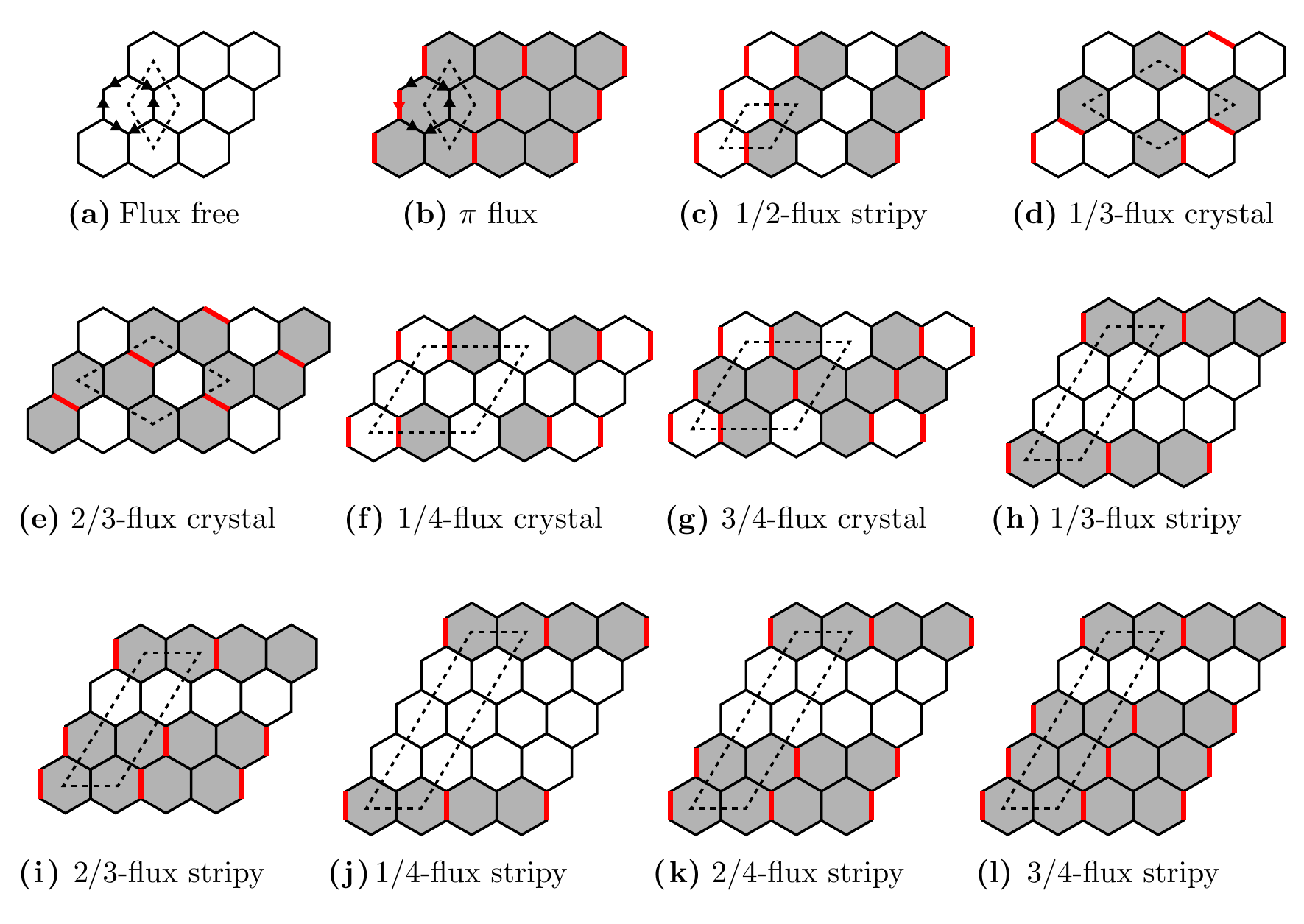}
    \caption{Same as Fig.~\ref{fig:fluxconfigs_sq}, but for the $\nu=3$ model.}
   \label{fig:fluxconfigs_hc}
\end{figure*}

\begin{figure}[!tb]
\includegraphics[width=.9\columnwidth,clip]{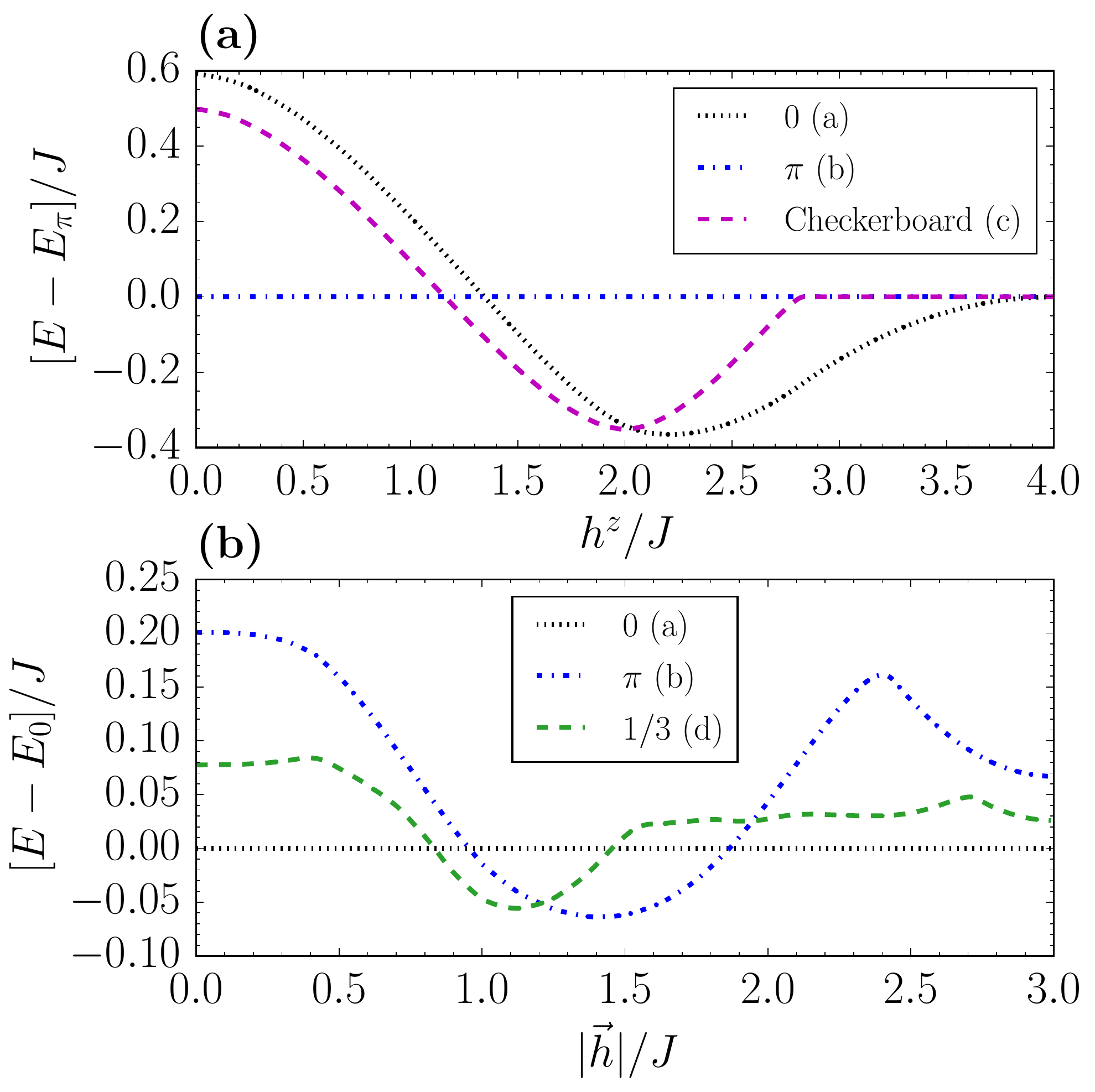}
\caption{
(a) Energy per unit cell in the respective flux sectors for the $\nu=2$ model (square lattice) relative to the $\pi$-flux sector energy.
The labels in the legend refer to the configurations displayed in Fig.~\ref{fig:fluxconfigs_sq}.
(b) As (a), but for the $\nu=3$ model (honeycomb lattice) relative to the flux-free sector's energy. The labels in the legend refer to the configurations displayed in Fig.~\ref{fig:fluxconfigs_hc}.
}
\label{fig:gs_energs_fields}
\end{figure}

\begin{figure}[!tb]
\includegraphics[width=.9\columnwidth,clip]{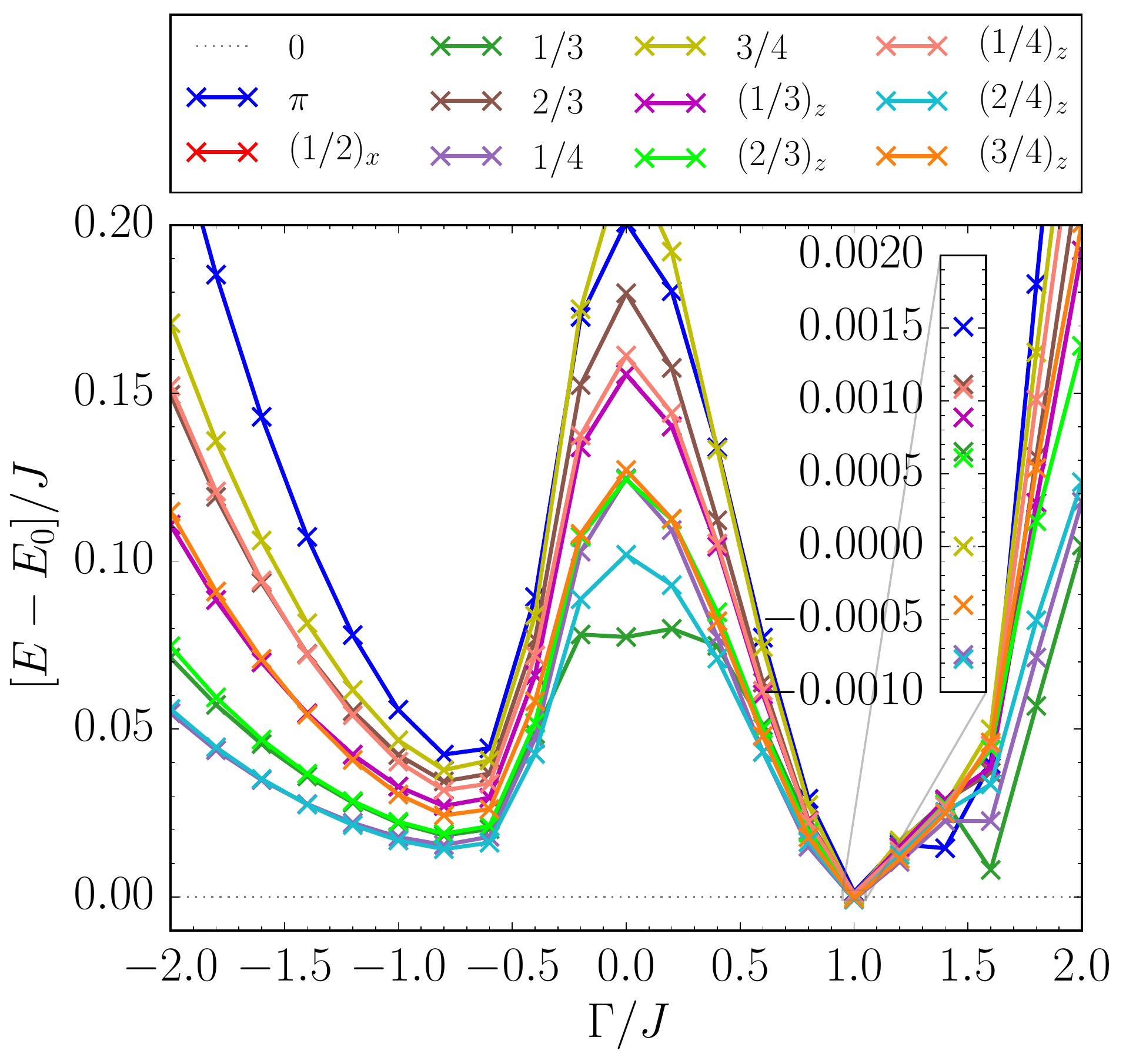}
\caption{
 Energies of $\tilde{\mathcal{H}}_J^{(3)} + \tilde{\mathcal{H}}_\Gamma^{(3)}$ in the various flux sectors, relative to the $0$-flux sector, as a function of $\Gamma/J$ obtained by diagonalization on a lattice with $36 \times 36$ unit cells. The labels $1/3$ etc. refer to the corresponding flux crystals displayed in Fig.~\ref{fig:fluxconfigs_hc}, and $(1/3)_a$ etc. denote stripy flux configurations with stripes perpendicular to $a$-type bonds. Inset: Energies at $\Gamma=J$.
}
\label{fig:gs_energs_gamma}
\end{figure}

We obtain the ground-state flux sector by fixing an appropriate gauge for the $\{u_{ij}\}$ and then comparing the ground-state energies of resulting free-fermion Hamiltonian in the respective flux sector, following the method by Kitaev \cite{kitaev06}.
The various flux patterns considered in the search for the ground-state flux sector are shown in Fig.~\ref{fig:fluxconfigs_sq} for the square lattice and Fig.~\ref{fig:fluxconfigs_hc} on the honeycomb lattice.
We note that in flux sectors with a total $\Ztwo$ flux per \emph{physical} unit cell, translation symmetry is implemented projectively, and thus any particular gauge-fixed configuration $\{u_{ij}\}$, giving rise to these respective flux sectors, enlarges the size of the Majorana unit cell~\cite{kitaev06, Zhang2019,Zhang2020}.

After gauge-fixing, the Majorana-fermion tight-binding Hamiltonian can be written in the form
\begin{equation} \label{eq:TBHam}
    \mathcal{H} = \frac{\iu}{4} \sum_{\substack{\alpha,i,s_1\\ \beta,j,s_2}} c_{\alpha,i,s_1} \mathcal{A}_{\alpha,i,s_1;\beta,j,s_2} c_{\beta,j,s_2}
\end{equation}
where $\alpha,\beta = x,y(z)$ indexes the Majorana flavours for the square (honeycomb) lattice.. We enumerate two-site unit cells on the bipartite square and honeycomb lattices by $i,j=1, \dots, N^2$, and let $s_1,s_2 = A,B$ denote the sublattice degrees of freedom. Note that the fermionic statistics imply that $\mathcal A$ is skew-symmetric.
As shown by Kitaev, the ground-state energy of $\mathcal{H}$ in \eqref{eq:TBHam} (per unit cell) is given by the sum over all negative eigenvalues $\varepsilon_\mu < 0$ of the matrix $\iu \mathcal{A}$,
\begin{equation}
    E/N^2 = \frac{1}{2 N^2} \sum_{\mu: \varepsilon_\mu <0} \varepsilon_\mu.
\end{equation}
Determining the ground-state energy of \eqref{eq:TBHam} thus amounts to diagonalizing a $4 N^2 \times 4 N^2$ ($6 N^2 \times 6 N^2$) matrix for the $\nu=2(3)$ models, respectively.

To minimize the computational time required, we first determine the ground-state energies in all flux sectors for smaller lattice sizes, and subsequently consider only the lowest-energy flux sectors to determine the respective phase boundaries on lattices with $N^2 = 48 \times 48$ unit cells.
The resulting energies (per unit cell) as a function of the applied Zeeman field are given in Fig.~\ref{fig:gs_energs_fields}.
For the off-diagonal $\Gamma$-type interaction on the honeycomb lattice, we find that the flux-free sector to be the optimal flux configuration every except at $\Gamma=J$, where many flux sectors are close in energy as displayed in Fig.~ \ref{fig:gs_energs_gamma}. Since at $\Gamma=J$ the flux-free sector's energy per unit cell $E_0 = -5.5798 J$, reliably determining the optimal ground-state flux sector requires a more systematic study with higher numerical accuracy.

Having diagonalized $\iu \mathcal A$ with an unitary transformations $U$ such that $U^\dagger \iu \mathcal A U = \mathrm{diag}(\epsilon_1, \dots)$, the magnetization $m^\alpha=1/(4N^2) \sum_i \langle \sigma^\alpha_i \rangle$ induced by a non-zero Zeeman field can be straightforwardly obtained.
The required fermionic bilinear expectation values can be computed by expanding the Majorana fermions in terms of the normal modes as $c_{\alpha,i,s} = \sum_{\mu: \varepsilon_\mu < 0} U_{\alpha,i,s;\mu} \gamma_\mu + U_{\alpha,i,s;\mu}^\ast \gamma_\mu^\dagger$.
Note that the sum extends over only the negative eigenvalues to avoid the redundancy of the Majorana spectrum \cite{kitaev06}.
The expectation values then read
\begin{equation}
    \langle \iu c^\alpha_{s,i} c^\beta_{s,i} \rangle = \sum_{\mu: \varepsilon_\mu < 0} \left[\iu U_{\alpha,i,s;\mu} U_{\mu;\beta,i,s}^\ast f_\mathrm{D}(\varepsilon_\mu) + \hc\right],
\end{equation}
where $f_\mathrm{D}$ denotes the Fermi Dirac distribution with $\lim_{T\to 0} f_\mathrm{D}(\varepsilon) = \Theta(-\varepsilon)$.



\bibliography{solfields_bib_v1}
\bibliographystyle{apsrev4-2}

\end{document}